\documentclass[12pt]{iopart}

\expandafter\let\csname equation*\endcsname\relax
\expandafter\let\csname endequation*\endcsname\relax

\usepackage{MyMainPreamble}
\usepackage{xfrac}
\usepackage{placeins}

\makeatletter
\newrobustcmd{\fixappendix}{%
 \patchcmd{\l@section}{1.5em}{7em}{}{}%
 \patchcmd{\l@subsection}{2.3em}{7em}{}{}%
}
\makeatother

\begin{document}
	\title[Entanglement Hamiltonians with Defects]{Entanglement Hamiltonians for Periodic Free Fermion Chains with Defects}
	\author{Gavin Rockwood}
	\address{Department of Physics and Astronomy, Rutgers University, 136 Frelinghuysen Rd, Piscataway, NJ 08854, US }

\begin{abstract}
	We study the half system entanglement Hamiltonians of the ground state of free fermion critical transverse field Ising model with periodic boundary conditions in the presence of defects. In general, we observe that these defects introduce non-local terms into the entanglement Hamiltonian, with the most significant being couplings across the defect that decay with distance. We also perform a limited entanglement Hamiltonian reconstruction using an ansatz and analyze how the fitted non-local couplings vary with defect strength.
\end{abstract}
\maketitle

\section{Introduction}
	Entanglement entropy is a commonly used and computable measure for quantifying entanglement in quantum systems. For a pure state $\ket{\psi}$ and a bipartition of the system into $A\cup \bar{A}$, the entanglement entropy is defined as 
	\begin{equation}
		S_A = -\Tr \rho_A\log\rho_A
	\end{equation}
	where $\rho_A = \Tr_{\bar{A}}\ket{\psi}\bra{\psi}$ is the reduced density matrix, making the entanglement entropy the von Neumann entropy of $\rho_A$. This is a widely studied measure of entanglement from conformal field theories and quantum field theories\cite{Pasquale_Calabrese_2004, H_Casini_2004, Kitaev_Preskill_2006}, classifying phase transitions \cite{Osterloh:2002sym,Turner_etal_2011}\footnote{See \cite{Li2022} for an overview of this topic.}, including measurement induced phase transitions \cite{Vasseur_etal_2019,Skinner_etal_2019,Li_etal_2018,Zabalo_etal_2020}\footnote{For a general review, see \cite{Dalmonte_etal}}.
	
	\begin{figure}[H]
		\centering
		\begin{tikzpicture}
			\tikzmath{
			coordinate \c;
			coordinate \a;
			coordinate \b;
			\AngleStep = 20;
			\seperation = 8;
			\radius = 2.5;
			\spinsize = 0.1;
			\majoranasep = 5;
			\majoranasize = 0.1;
			\JOffsetSpins = 1.0;
			\JOffsetMajoranas = 1.0;
			\gWidth = 2;
			\JWidth = 2;
			\sigmazoffset = 4;
			\verticalshift = -2-\radius;
			\aa = -10*\spinsize+\verticalshift;
			for \x in {\AngleStep,2*\AngleStep,...,180-\AngleStep} {
				{\draw[line width = \JWidth, color = JBond] (\x: {\JOffsetSpins*\radius}) -- (\x+\AngleStep: {\JOffsetSpins*\radius});};
				{\draw[line width = \JWidth, color = JBond] (\x+180: {\JOffsetSpins*\radius}) -- (\x+\AngleStep+180: {\JOffsetSpins*\radius});};
			};
			{\draw[line width = \JWidth, color = black, dashed] (0: {\JOffsetSpins*\radius}) -- (\AngleStep: {\JOffsetSpins*\radius});};
			{\draw[line width = \JWidth, color = black, dashed] (180: {\JOffsetSpins*\radius}) -- (180+\AngleStep: {\JOffsetSpins*\radius});};
			for \x in {0,\AngleStep,...,360} {
				{\draw[line width=\gWidth, gBond] (\x: {\radius}) -- (\x: {1.2*\radius});};
			};
			for \x in {0,\AngleStep,...,360} {
				\c = (\x: {\radius});
				{\node[rectangle, draw, rotate = \x, fill = black, minimum width=\spinsize*1cm,minimum height=\spinsize*2cm] at (\c) {};};
			};
			for \x in {\AngleStep,2*\AngleStep,...,180-\AngleStep} {
				\a = (\seperation, 0) + (\x+\majoranasep: {\JOffsetMajoranas*\radius});
				\b = (\seperation, 0) + (\x-\majoranasep+\AngleStep: {\JOffsetMajoranas*\radius});
				{\draw[line width=\JWidth, JBond] (\a) -- (\b);};
				\a = (\seperation, 0) + (\x+\majoranasep+180: {\JOffsetMajoranas*\radius});
				\b = (\seperation, 0) + (\x-\majoranasep+\AngleStep+180: {\JOffsetMajoranas*\radius});
				{\draw[line width=\JWidth, JBond] (\a) -- (\b);};
			};
			\a = (\seperation, 0) + (\majoranasep: {\JOffsetMajoranas*\radius});
			\b = (\seperation, 0) + (0-\majoranasep+\AngleStep: {\JOffsetMajoranas*\radius});
			{\draw[line width=\JWidth, black, dashed] (\a) -- (\b);};
			\a = (\seperation, 0) + (\majoranasep+180: {\JOffsetMajoranas*\radius});
			\b = (\seperation, 0) + (0-\majoranasep+\AngleStep+180: {\JOffsetMajoranas*\radius});
			{\draw[line width=\JWidth, black, dashed] (\a) -- (\b);};
			for \x in {0,\AngleStep,...,360} {
				\a = (\seperation, 0) + (\x-\majoranasep: {\JOffsetMajoranas*\radius});
				\b = (\seperation, 0) + (\x+\majoranasep: {\JOffsetMajoranas*\radius});
				{\draw[line width=\gWidth, gBond] (\a) -- (\b);};
			};
			for \x in {0,\AngleStep,...,360} {
				\c = (\seperation, 0) + (\x+\majoranasep: {\radius});
				{ \fill (\c) circle [radius={\majoranasize} ]; };
				\c = (\seperation, 0) + (\x-\majoranasep: {\radius});
				{\node[rectangle, draw, rotate = \x-\majoranasep, fill = black, minimum width=\majoranasize*1.41cm,minimum height=\majoranasize*1.41cm, inner sep=2pt] at (\c) {};};
			};
			{
			\draw[line width = \JWidth, color = JBond] (-\radius/2, \verticalshift) -- (\radius/2, \verticalshift);
			\draw[line width = \gWidth, color = gBond] (-\radius/2, \verticalshift) -- (-\radius/2, \verticalshift+\sigmazoffset*\spinsize);
			\node[rectangle, draw, fill = black, minimum width=\spinsize*2cm,minimum height=\spinsize*1cm] at (-\radius/2, \verticalshift) {};
			\node[rectangle, draw, fill = black, minimum width=\spinsize*2cm,minimum height=\spinsize*1cm] at (\radius/2, \verticalshift) {};
			\node at (-\radius/2, \verticalshift+1.5*\sigmazoffset*\spinsize) {$g_j\sigma^z_j$};
			\node at (0, \verticalshift+\spinsize*\sigmazoffset) {$J_j\sigma^x_j\sigma_{j+1}^x$};
			};
			{
			\draw[line width = \JWidth, color = DualityBond] (-\radius/2, \aa).. controls (0, -0.5+\aa)..(\radius/2, \aa);
			\node[rectangle, draw, fill = black, minimum width=\spinsize*2cm,minimum height=\spinsize*1cm] at (-\radius/2, \aa) {};
			\node[rectangle, draw, fill = black, minimum width=\spinsize*2cm,minimum height=\spinsize*1cm] at (\radius/2, \aa) {};
			\node at (0, \aa+\spinsize) {$b_j\sigma^x_j\sigma_{j+1}^y$};
			};
			{
			\draw[line width = \JWidth, color = JBond] (\seperation-\radius/2+\radius/4,\verticalshift) -- (\seperation+\radius/2-\radius/4,\verticalshift);
			\draw[line width = \gWidth, color = gBond] (\seperation-\radius/2-\radius/4,\verticalshift) -- (\seperation-\radius/2+\radius/4,\verticalshift);
			\fill (\seperation-\radius/2-\radius/4,\verticalshift) circle [radius={\majoranasize} ];
			\node[rectangle, draw, fill = black, minimum width=\majoranasize*1.41cm,minimum height=\majoranasize*1.41cm] at (\seperation-\radius/2+\radius/4,\verticalshift) {};
			\fill (\seperation+\radius/2-\radius/4,\verticalshift) circle [radius={\majoranasize} ];
			\node[rectangle, draw, fill = black, minimum width=\majoranasize*1.41cm,minimum height=\majoranasize*1.41cm] at (\seperation+\radius/2+\radius/4,\verticalshift) {};
			\node at (\seperation-\radius*0.75, \verticalshift+\sigmazoffset*\spinsize) {$g_j\gamma_{2j}\gamma_{2j+1}$};
			\node at (\seperation+\radius*0.25, \verticalshift+\sigmazoffset*\spinsize) {$J_j\gamma_{2j+1}\gamma_{2j+2}$};
			};
			{
			\draw[line width = \JWidth, color = DualityBond] (\seperation-\radius/2+\radius/4,\aa) .. controls (\seperation+\radius/4,-0.5+\aa) .. (\seperation+\radius/2+\radius/4,\aa);
			\fill (\seperation-\radius/2-\radius/4,\aa) circle [radius={\majoranasize} ];
			\node[rectangle, draw, fill = black, minimum width=\majoranasize*1.41cm,minimum height=\majoranasize*1.41cm] at (\seperation-\radius/2+\radius/4,\aa) {};
			\fill (\seperation+\radius/2-\radius/4,\aa) circle [radius={\majoranasize} ];
			\node[rectangle, draw, fill = black, minimum width=\majoranasize*1.41cm,minimum height=\majoranasize*1.41cm] at (\seperation+\radius/2+\radius/4,\aa) {};
			\node at (\seperation+\radius/4, \aa+\sigmazoffset*\spinsize) {$b_j\gamma_{2j+1}\gamma_{2j+3}$};
			};
			}
		\end{tikzpicture}\caption{The general geometry before (on left) and after (on right) the Jordan Wigner transform. Post JW transform, we end up with a Majorana hopping model where the duality defect (in gold) leads to a term that skips a Majorana site. Because we index from $0$ to $N-1$, we call the circle ``even'' whereas the square sites are considered ``odd''.}\label{fig:GeneralSetup}
	\end{figure}

	The entanglement entropy, however, distills all the information of $\rho_A$ down into a single number and does not fully capture all there is to know about the entanglement structure of the state. In this paper, we are focusing on the entanglement Hamiltonian, $\mathcal{K}_A$, an object that contains all of the entanglement information about $\rho_A$, and can be used to both extract qualitative and quantitative properties of the entanglement structure \cite{H_Casini_2004, Cardy:2016fqc,LiHaldane_2008, Wong:2013gua,Eisler_2024}. Defined via
	\begin{equation}\label{eq:DefofModHam}
		\rho_A = \frac{1}{Z}e^{-\mathcal{K}_A},
	\end{equation}
	we can interpret the reduced density matrix as a thermal state of the entanglement Hamiltonian with $\beta = 1$. The spectrum of $\mathcal{K}_A$ ($\curly{\epsilon_i}$), called the entanglement spectrum, is directly related to the Schmidt spectrum ($\{\lambda_i\}$) by $\epsilon_i+\epsilon_{0} = -\log\lambda^2_i$, where $\epsilon_0 = -\log Z$\footnote{It is also common to see the entanglement Hamiltonian written as $\rho_A = e^{-\mathcal{K}_A}$, in which case $\epsilon_i = -\log\lambda_i^2$. In this paper we pull out the vacuum energy and write it as the $1/Z$ factor in equation \ref{eq:DefofModHam}.}. It is the structure of $\mathcal{K}_A$ that we are exploring in this paper, an exploration that is largely motivated by the behavior of the entanglement spectrum compared to entanglement entropy for a certain collection of states. In particular, our collection of states will be the ground states of a free fermion chain with defects where the entanglement entropy is largely independent of the defect value whereas the entanglement spectrum is not.
	
	This paper in part builds on the ideas in \cite{RoySaleur2022EE,Roy2022,RogersonPollmannRoy2022}, where the focus is on the periodic transverse field Ising model (equation \ref{eq:TFI_Ham}), in the critical phase, using the language of free fermions. Our starting Hamiltonian is 
	\begin{equation}\label{eq:TFI_Ham}
		H = -\frac{1}{2}\sum_{i\in\Z_N}\para{\sigma_i^x\sigma_{i+1}^x + \sigma_i^z} = -\frac{1}{2}\sum_{i\in\Z_N}\para{\gamma_{2i+1}\gamma_{2i+2}+\gamma_{2i}\gamma_{2i+1}},
	\end{equation}
	where $\gamma_m$ are the Majorana operators with $\curly{\gamma_m,\gamma_n} = 2\delta_{mn}$ (see figure \ref{fig:GeneralSetup}). From here, we will introduce energy and duality defects, defined as: 
	\begin{equation}\label{eq:EnergyDefect}
	H_{\text{energy}, i} = -\frac{1}{2}(J_i^*-1)\sigma_{i}^x\sigma_{i+1}^x = 
	-\frac{1}{2}(J_i^*-1)\gamma_{2i+1}\gamma_{2i+2},
	\end{equation}
	\begin{equation}\label{eq:DualityDefect}
	\begin{aligned}
	H_{\text{duality}, i} &= -\frac{1}{2}\para{b_i\sigma_i^x\sigma_{i+1}^y - J_i\sigma_i^x\sigma_{i+1}^x - g_{i+1}\sigma_{i+1}^z}\\
	&=-\frac{1}{2}\para{b_i\gamma_{2i+1}\gamma_{2i+3} - J_i\gamma_{2i+1}\gamma_{2i+2} - g_{i+1}\gamma_{2i+3}\gamma_{2i+4}}.
	\end{aligned}
	\end{equation}
	Adding an energy defect at site $i$ changes the coupling from $1$ to $J_i^*$, while adding a duality defect at site $i$ removes the $\sigma_i^x\sigma_{i+1}^x$ and $\sigma_{i+1}^z$ terms, and replaces them with $b_i\sigma_i^x\sigma_{i+1}^y$. In the Majorana language, the duality defect decouples one of the Majorana modes from the system. With the definition we have chosen for the duality defect, we decouple an even flavored Majorana (circles in figure \ref{fig:GeneralSetup}).\footnote{There are two possible definitions of the duality defect, one with $\sigma_i^x\sigma_{i+1}^y$ (which we are using) and $\sigma_i^y\sigma_{i+1}^x$. The latter changes remove $\sigma_i^z$ instead of $\sigma_{i+1}^z$. The end result is similar, except the flavor of decoupled Majorana is odd instead of even.}  

	\begin{figure}[H]
		\centering
		\begin{minipage}[c]{0.45\textwidth}
			\subcaptionbox{\label{fig:EntanglementEntropyIntro}}{\includegraphics[scale = 0.7]{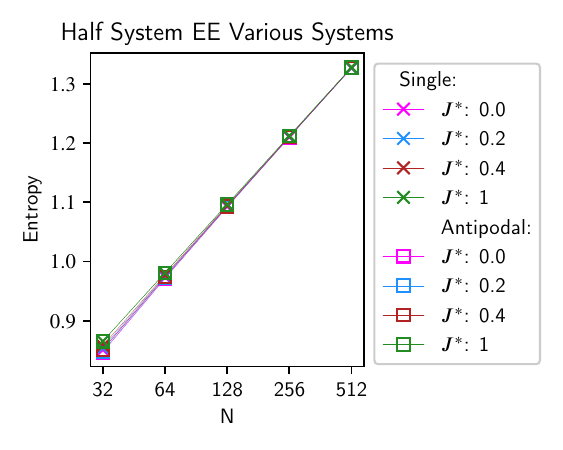}}
		\end{minipage}
		\begin{minipage}[c]{0.45\textwidth}
			\subcaptionbox{\label{fig:EntanglementSpecIntro}}{\includegraphics[scale = 0.7]{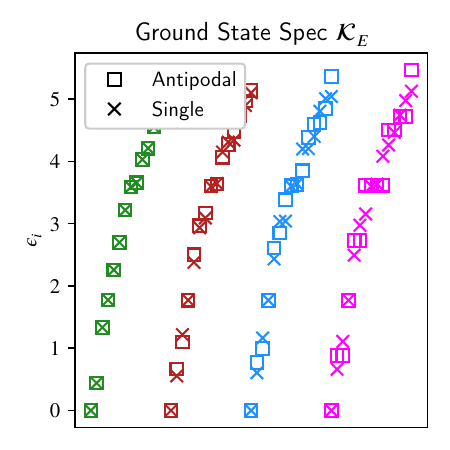}}
		\end{minipage}
		\caption{\ref{sub@fig:EntanglementEntropyIntro}) Entanglement entropy for various single and antipodal energy defects as a function of system size $N$. \ref{sub@fig:EntanglementSpecIntro}) The entanglement spectra for systems of size $N=8129$ for various single and antipodal defects. The colors correspond to the defect strength, and the offset within the columns is just for visual clarity and is a function of the eigenvalue index. The key takeaway is that the entanglement entropy converges for large system sizes regardless of defect strength. However, the entanglement spectrum distinguishes the defects even for very large systems.}\label{fig:MainMotivationFig} 
	\end{figure}

	The motivation for this paper can be summarized in figure \ref{fig:MainMotivationFig}, where we have calculated the entanglement entropy and entanglement spectra where the subsystem $A$ is half the system and contained at the center of $A$ is an energy defect (Single case) and where $\bar{A}$ also contains a defect opposite the one in $A$ (Antipodal case).\footnote{For a system of size $N$ and where $A=[0,N/2-1]$, the single case is a defect at $i = N/4-1$ (see fig \ref{fig:SingleDefectSetup}) and the antipodal has defects at $N/4-1$ and $3N/4-1$ (see fig \ref{fig:AntipodalDefectSetup}).} As the system size $N$ grows, the lattice effects become less important, and the entanglement entropies of all these configurations converge. The entanglement entropy is effectively ignorant of the defect; however, the defect has an impact on the entanglement spectra (figure \ref{fig:EntanglementSpecIntro}). This means that $\mathcal{K}_A$ must also have knowledge of these defects in the large system limit, the goal of this paper is to see how this dependence manifests using high precision free fermion techniques. One feature we expect to see in the entanglement Hamiltonian has non-local interactions across the defect, a feature that can be shown to be expected by looking at the fermionic logarithmic negativity (see \ref{sec:FLNArgument}). We are also interested in what we can predict about the continuum limit CFT entanglement Hamiltonian. We can use the non-local terms in $\mathcal{K}_A$ to predict what the non-local structure (if any) the CFT entanglement Hamiltonian should have, and, coupling this with the Hamiltonian reconstruction technique discussed in \cite{PhysRevB.99.235109}, we can check our ansatz and study how the strength of our predicted couplings vary with respect to the defect strengths.   
	
	This is not the first work to discuss high precision calculations of entanglement Hamiltonians using free fermion methods. For small critical systems, see \cite{Eisler_Peschel_2017,Eisler_Peschel_2018}; for two disjoint intervals, see \cite{Eisler_2022}; for non-critical spin chains, see \cite{Eisler_2020}; for quenches, see \cite{Di_Giulio_2019, rottoli2024entanglementhamiltoniansquasiparticlepicture}.

\section{Free Fermion Techniques}\label{sec:FFTechniques}
	The entanglement Hamiltonian of a free complex fermion system can be written as $\mathcal{K}_A = \sum_{i,j\in A}(K_A^\mathbb{C})_{ij}c_i^\dagger c_j$ where the matrix $K_A^\mathbb{C}$ is given by the equation  
	\begin{equation}\label{eq:ComplexFermionK}
		K^{\mathbb{C}}_A = -\log\brac{2(G_A^\mathbb{C})^{-1} - \mathbbm{1}}, 
	\end{equation}
	where $G_A^\mathbb{C}$ is the restriction of the two point correlation matrix $G^\mathbb{C}_{ij} = \braket{\para{\vec{c}\cdot(\vec{c})^\dagger}_{ij}}$, $\para{\vec{c}}^T = \para{c_0^\dagger, \eli, c_{N-1}^\dagger, c_0,\eli, c_{N-1}}$, to indices in $A$\cite{Ingo_Peschel_2003}. The $\mathbb{C}$ superscript denotes the complex fermion basis. The extra factor of $2$ in equation \ref{eq:ComplexFermionK} compared to reference \cite{Ingo_Peschel_2003} comes from the fact we are using the full two point matrix, not just $\braket{c_i^\dagger c_j}$.  Since we will be transitioning to Majorana fermions, $\{\gamma_m\}$, from complex fermions,
	\begin{equation}
		c_i^\dagger = \gamma_{2i}+\i\gamma_{2i+1},\hspace{1cm} c_i = \gamma_{2i}-\i\gamma_{2i+1},
	\end{equation}
	we can denote the correlation matrix for the Majorana fermions as $G_{mn}^M=\braket{\gamma_m\gamma_n}$. However, because we will be staying in this basis from here on out we will drop this superscript. 

	Following \cite{Latorre:2003kg} to obtain $G_A$, we start with a Hamiltonian for $2N$ Majorana fermions written as $\mathcal{H} = \tfrac{1}{2}\sum_{mn}^{2N}H_{mn}\gamma_m\gamma_n$, where $H$ is an antisymmetric matrix with purely imaginary entries (making it Hermitian). We then perform real Schur decomposition on $A = -2\i H$ and obtain $A = UBU^{\dagger}$. Because we are doing real Schur decomposition, $B$ will be a matrix of the form
	\begin{equation}
		B = \bigoplus_{i=0}^{N-1}\begin{pmatrix} 0 & -\epsilon_i\\ \epsilon_i & 0 \end{pmatrix}.
	\end{equation}
	The ground state two point correlation matrix is then given by $G = \mathbbm{1}+\i\Gamma_A$ where $\Gamma_A = U\Gamma_BU^{\dagger}$ with 
	\begin{equation}
		\Gamma_B = \bigoplus_{i=0}^{N-1}\begin{pmatrix}0 & -1 \\ 1 & 0\end{pmatrix}.
	\end{equation}

\begin{figure}[H]
		\centering\hspace{-4cm}
		\begin{minipage}[c]{0.8\textwidth}
			\subcaptionbox{\label{fig:NoDefectNN}}{\includegraphics[scale = 0.7]{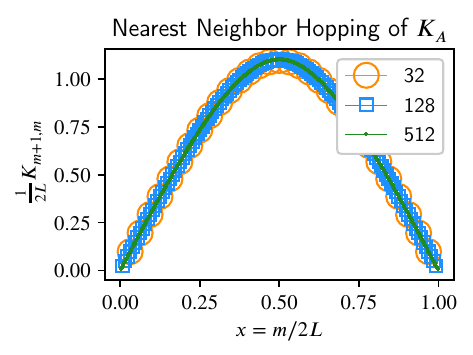}}
			\subcaptionbox{\label{fig:NoDefectSymmetricHopping}}{\includegraphics[scale = 0.7]{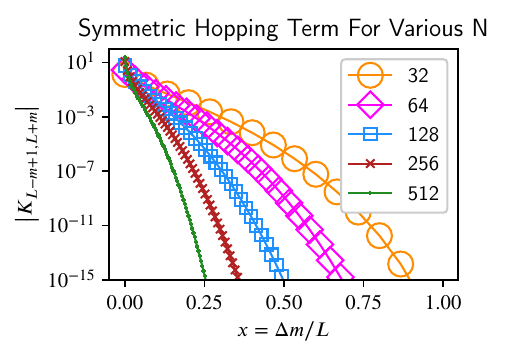}}
		\end{minipage}
		\begin{minipage}[c]{0.1\textwidth}
			\subcaptionbox{\label{fig:NoDefect_K}}{\includegraphics[scale = 0.7]{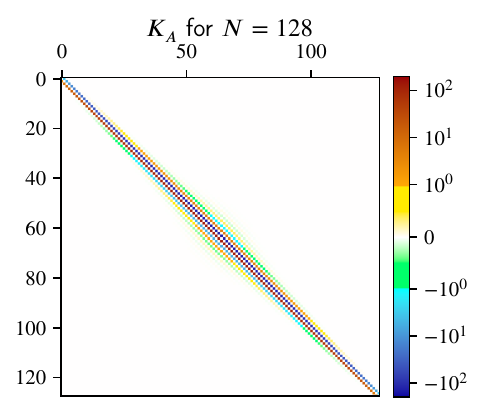}}
		\end{minipage}
		\caption{All the figures are looking at elements of $K$ for no defect. \ref{sub@fig:NoDefectNN}) Plot of Nearest neighbor couplings of $K$.  \ref{sub@fig:NoDefectSymmetricHopping}) Plot of the symmetric hopping terms / anti-diagonal entries of $K_A$ (for example, coupling the first and last sites of the subsystem). These terms are the largest non-local terms in the entanglement Hamiltonian and will become prominent when defects are introduced. \ref{sub@fig:NoDefect_K}) Plot of the entries of $K_A$. The scale is symmetric log where the linear threshold is $\pm1$.}\label{fig:SomePlotsForNoDefect1}
	\end{figure}

	The reduced correlation matrix is then given by the restriction of $G$ to some sub-region $A$.  To obtain the entanglement Hamiltonian matrix $K_A$, we calculate $K_A = -\log\brac{2\para{G_A}^{-1} - \mathbbm{1}}$. Because $\tfrac{1}{2}G_A$ has eigenvalues that are increasingly close to $0$ and $1$, evaluating this equation requires high numerical precision, which scales with system size. This was also observed in \cite{Eisler_Peschel_2017, Eisler_Peschel_2018} where subsystems of size $L=40$ were analyzed. In this paper, we used the Python package ``mpmath"\cite{mpmath} to push the system size up to $N=512$. Most of these calculations were performed with a decimal precision of at least $1.5N$ and up to $4N$. This precision ultimately depends on the choice of $A$ and the parameters of the Hamiltonian. It should also be noted that when looking at $K_A$, we will be keeping it in the Majorana basis such that $\mathcal{K}_A = \sum_{m,n\in A}(K_A)_{mn}\gamma_m\gamma_n$. In this basis, $K_A$ will be purely imaginary and of dimension $2L$ for a subsystem of $L$ complex fermions. \textbf{\textit{For all of our plots of Majorana $K_A$ and its elements, it is implied that we are plotting $\im (K_A)$}.} We will also consistently use $m$ to index the Majorana sites in $A$, with $m\in[0, N-1]$ for a half-system subsystem for a chain of $N\in4\mathbb{N}$ complex fermions. Sticking to multiples of $4$ for the system size allows us to have centered energy defects within a half system. We will also occasionally denote $K_A^{J^*}$ as the entanglement Hamiltonian for a choice of defect $J^*$ when comparing multiple defect strengths. We will also drop the subsystem subscript if it is not needed. Results for a no defect critical, periodic TFI chain are shown in figure \ref{fig:SomePlotsForNoDefect1} as a baseline for comparison. The important takeaway from figure \ref{fig:SomePlotsForNoDefect1} is that the entanglement Hamiltonian, while not technically local ($K_A$ is checkerboard dense with only odd-even couplings), is dominated by local interactions. These Hamiltonians are in agreement with the analytic results from \cite{Eisler_Peschel_2017}. 

	It is also worth noting that in \cite{Eisler_Peschel_2017}, they show that for an infinite spin chain, the lattice entanglement Hamiltonian coupling matrix for a subsystem of size $L$ is given by a power series of the CFT prediction $K^\text{CFT} = 4LT$ where
	\begin{equation}\label{eq:T_Inf}
		T = \sum_{n=0}^{2L-2}\frac{n+1}{2L}\para{1-\frac{n+1}{2L}}.
	\end{equation}
	From this, they show that $K$ is given by
	\begin{equation}\label{eq:K_Lat}
		K^{\text{lat}} = 4L\sum_{m=0}^\infty \alpha_m\beta_m T^{2m+1},
	\end{equation}
	where
	\begin{equation}
		\begin{aligned}
			\alpha_m  = \frac{1}{\sqrt{\pi}}\frac{\Gamma(m+\tfrac{1}{2})}{\Gamma(m+1)},\hspace{1cm} \beta_m = \sqrt{\pi}2^{2m}\frac{\Gamma(2m+\tfrac{1}{2})}{\Gamma(2m+2)}.
		\end{aligned}
	\end{equation}
	They argue that in the continuum limit, only the $m=0$ term survives, yielding the CFT prediction. For the periodic system, this also works if we replace the $n$ dependence in equation \ref{eq:T_Inf} with $\sin\brac{\frac{\pi(n+1)}{2L}}$. An important consequence of this relationship is that $K^\text{lat}$ and $K^\text{CFT}$ must commute, and thus must share the same eigenvectors. If we work on the assumption that this should at least roughly hold true for systems with defects, we can use the scaling behavior of $K$ to make predictions about the form of the entanglement Hamiltonian in the continuum limit and even perform a Hamiltonian reconstruction technique (using an ansatz for the required operators), to make assertions about the exact form of $K^\text{CFT}$. 

	\subsection{Entanglement Hamiltonian Reconstruction}
		Following the method outlined in \cite{PhysRevB.99.235109}, we will use our results for $K$ and its scaling behavior to try and make predictions about what the continuum limit entanglement Hamiltonian might look like. This technique involves taking the entanglement ground state $\ket{\xi}$ and a set of operators $\curly{L_\alpha}$ and then determining a set of weights $\curly{\omega_i}$ such that the approximate entanglement Hamiltonian is given by 
		\begin{equation}
			\tilde{\mathcal{K}} = \sum_{\alpha}\omega_\alpha L_\alpha.
		\end{equation}
		This approximation is built out of a selection of operators that we expect to be present in the continuum limit. In \cite{PhysRevB.99.235109}, the authors were able to predict the known CFT entanglement Hamiltonian for the XY chain where the operators $L_i = S_i^xS^x_{i+1}+S_i^yS^y_{i+1}$. In this paper, we are interested in predicting the form of the CFT entanglement Hamiltonian in the presence of defects. 
		In order to calculate $\curly{\omega_i}$, we start by constructing the matrix
		\begin{equation}
			G_{ab} = \braket{\xi|L_aL_b|\xi} - \braket{\xi|L_a|\xi}\braket{\xi|L_b|\xi}.
		\end{equation}
		$G$ is a positive semidefinite matrix, and the desired weights are given by the lowest eigenvalue of $G$, denoted $g_0$. This value, $g_0$, is the energy fluctuation of $\tilde{\mathcal{K}}$. In the limit of $g_0\to0$, our state $\ket{\xi}$ is an exact eigenstate of $\tilde{\mathcal{K}}$. In \cite{PhysRevB.99.235109}, they achieved $g_0$ values on the order of $10^{-8}$ for $L=32$, where $L$ is the subsystem size. We aim to match or surpass this for larger system sizes. The presented method does run into convergence issues for large system sizes and operator counts, this is because we only have one constraint on our reconstruction procedure, the entanglement ground state. As we introduce extra couplings, we get highly degenerate possible Hamiltonians, making it hard to make a reasonable prediction.  Due to this, we need to have a good guess for what the operators should be and thus, we will only be applying this to the centered single and antipodal defects, systems where we have been able to find effective operators. In these cases, our operators will be 
		\begin{equation}\label{eq:DefOfLs}
			\begin{aligned}
				L_1 &= \sum_{n=0}^{2L-2}\sin\brac{\frac{\pi(n+1)}{2L}}H_{n,n+1}\gamma_n\gamma_{n+1},\\
				L_2 &= \sum_{n=0}^{L-1}(-1)^{n}\sin\brac{\frac{\pi(n+1)}{2L}}\gamma_n\gamma_{2L-n-1}.
			\end{aligned}
		\end{equation}
		$L_1$ is the standard CFT prediction for the Ising model (up to a factor of $4L$ left out for stability reasons), and is just a local rescaling of the stress energy tensor. In the case with no defects, this is the known CFT result and can be reconstructed by fitting each individual coupling constant using this technique. It is also known that for the massless Dirac CFT on an infinite line with a defect centered within a subsystem, the resulting entanglement Hamiltonian takes the form of a locally rescaling of the stress-energy tensor\cite{Mintchev_Tonni_2021}. In this case, there are no non-local components and the effect of the defect manifests entirely in $T_{00}$. Because we are focusing on periodic systems\footnote{Actually, we will see this is a consequence of the chain being a finite system.}, we will need to introduce a non-local component to our predictions, hence the operator $L_2$ which couples symmetrically across the defect. While this choice of couplings and envelope seems arbitrary, the motivation will become clear when looking at the dominant non-local terms coming from the single centered defect. The envelope was initially chosen as it seemed a natural choice given the known envelope for the local interactions.
		
		It is also important to note that when solving for the weights $\omega_i$, the eigenvalue solvers will return a normalized vector, and thus we will need to rescale the weights. This factor will be the $4L/\omega_1$, where $4L$ is the factor that we left out when defining operator $L_1$ and $1/\omega_1$ is included to fix the overall factor of $-1$. Rescaling by $1/\omega_{1}$ does not change the value by much as $1-|\omega_1|$ is $\mathcal{O}(10^{-4})$ or smaller. All plots and discussions of the weights will assume that this rescaling has been done and thus we will not discuss $\omega_1$ as it will be fixed to $4L$, leaving $\omega_2$ to be the sole parameter of interest in the reconstruction discussions. 

		The main body of this text will be focusing on the centered single and antipodal defects as these are the geometries where we have been able to provide an ansatz for the continuum limit entanglement Hamiltonian. The appendix will contain discussions for off-centered defects and the duality defect.
		
\section{Antipodal Defects}\label{sec:AntipodalDefect}
	\begin{figure}[h]
		\centering
		\begin{tikzpicture}[scale = 0.65]
			\tikzmath{
			coordinate \c;
			coordinate \a;
			coordinate \b;
			\AngleStep = 22.5;
			\radius = 2.5;
			\spinsize = 0.1;
			\majoranasep = 5;
			\majoranasize = 0.1;
			\JOffsetSpins = 1.0;
			\JOffsetMajoranas = 1.0;
			\gWidth = 2;
			\JWidth = 2;
			\sigmazoffset = 4;
			\verticalshift = -2-\radius;
			\a = (\AngleStep-\majoranasep/2-\majoranasize*1.41cm-\AngleStep/4: {\JOffsetMajoranas*\radius});
			\b = (\AngleStep-\majoranasep/2+180-\majoranasize*1.41cm-\AngleStep/4: {\JOffsetMajoranas*\radius+0.5});
			{\draw[line width = 6, color = black!40] (\a) arc[start angle = \AngleStep-\majoranasep/2-\majoranasize*1.41cm-\AngleStep/4, end angle = \AngleStep-\majoranasep/2+180-\majoranasize*1.41cm-\AngleStep/4, radius = \radius];};
			for \x in {\AngleStep,2*\AngleStep,...,180} {
				\a = (\x+\majoranasep: {\JOffsetMajoranas*\radius});
				\b = (\x-\majoranasep+\AngleStep: {\JOffsetMajoranas*\radius});
				{\draw[line width=\JWidth, JBond] (\a) -- (\b);};
				\a = (\x+\majoranasep+180: {\JOffsetMajoranas*\radius});
				\b = (\x-\majoranasep+\AngleStep+180: {\JOffsetMajoranas*\radius});
				{\draw[line width=\JWidth, JBond] (\a) -- (\b);};
			};
			\a = (\majoranasep+180: {\JOffsetMajoranas*\radius});
			\b = (0-\majoranasep+\AngleStep+180: {\JOffsetMajoranas*\radius});
			for \x in {0,\AngleStep,...,360} {
				\a = (\x-\majoranasep: {\JOffsetMajoranas*\radius});
				\b = (\x+\majoranasep: {\JOffsetMajoranas*\radius});
				{\draw[line width=\gWidth, gBond] (\a) -- (\b);};
			};
			for \x in {0,\AngleStep,...,360} {
				\c = (\x+\majoranasep: {\radius});
				{ \fill (\c) circle [radius={\majoranasize} ]; };
				\c = (\x-\majoranasep: {\radius});
				{\node[rectangle, draw, rotate = \x-\majoranasep, fill = black, minimum width=\majoranasize*1.41cm,minimum height=\majoranasize*1.41cm, inner sep=2pt] at (\c) {};};
			};
			\a = (\AngleStep-\majoranasep/2-\majoranasize*1.41cm-\AngleStep/4: {\JOffsetMajoranas*\radius+0.5});
			\b = (\AngleStep-\majoranasep/2+180-\majoranasize*1.41cm-\AngleStep/4: {\JOffsetMajoranas*\radius+0.5});
			\a = (90+\AngleStep-\majoranasep/2-\majoranasize*1.41cm-\AngleStep/4: {\JOffsetMajoranas*\radius+0.5});
			\b = (90+\AngleStep-\majoranasep/2-\majoranasize*1.41cm-\AngleStep/4: {\JOffsetMajoranas*\radius-0.5});
			{\draw[thick, decoration = zigzag, decorate] (\a) -- (\b);};
			{\node[below] at (\b) {$J^*$};};
			\a = (-90+\AngleStep-\majoranasep/2-\majoranasize*1.41cm-\AngleStep/4: {\JOffsetMajoranas*\radius+0.5});
			\b = (-90+\AngleStep-\majoranasep/2-\majoranasize*1.41cm-\AngleStep/4: {\JOffsetMajoranas*\radius-0.5});
			{\draw[thick, decoration = zigzag, decorate] (\a) -- (\b);};
			{\node[above] at (\b) {$J^*$};};
			\a = (45+\AngleStep-\majoranasep/2-\majoranasize*1.41cm-\AngleStep/4: {\JOffsetMajoranas*\radius+1});
			{\node[below] at (\a) {$A$};};
			}
		\end{tikzpicture}
		\caption{Diagram for the antipodal defects.}\label{fig:AntipodalDefectSetup}
	\end{figure}
\begin{figure}[h]
		\centering\hspace{-3cm}
		\begin{minipage}[c]{0.8\textwidth}
			\subcaptionbox{\label{fig:AntipodalDefect_NN_Ns}}{\includegraphics[scale = 0.7]{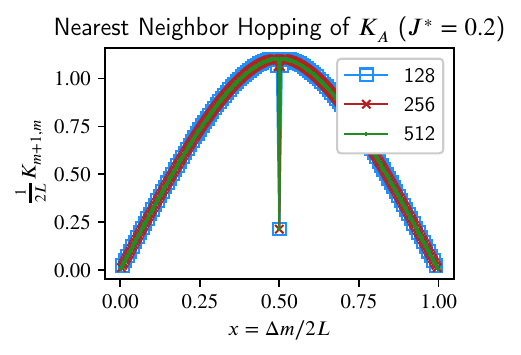}}
			\subcaptionbox{\label{fig:AntipodalDefect_NN_Defects}}{\includegraphics[scale = 0.7]{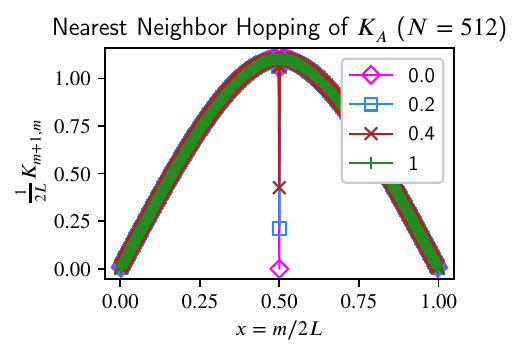}}
		\end{minipage}
		\begin{minipage}[c]{0.1\textwidth}
			\subcaptionbox{\label{fig:AntipodalDefect_K}}{\includegraphics[scale = 0.7]{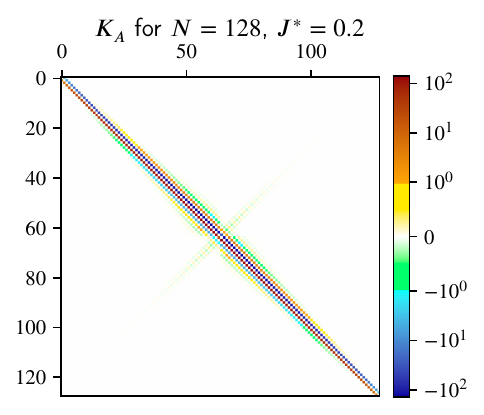}}
		\end{minipage}
		\begin{minipage}[c]{0.45\textwidth}
			\subcaptionbox{\label{fig:AntipodalDefect_NN_Dif}}{\includegraphics[scale = 0.7]{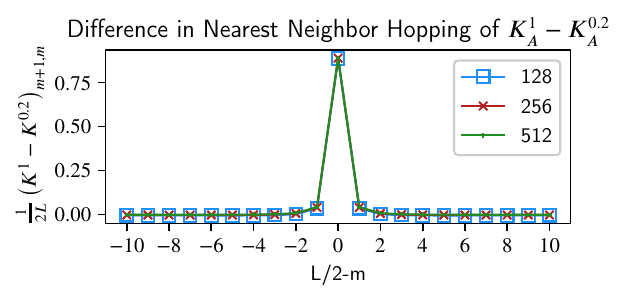}}
		\end{minipage}
		\caption{\ref{sub@fig:AntipodalDefect_NN_Ns}) Plot of nearest neighbor interactions for various system sizes for $J^* = 0.2$. \ref{sub@fig:AntipodalDefect_NN_Defects}) Plot of nearest neighbor interactions for various defects for $N=512$. \ref{sub@fig:AntipodalDefect_K}) The full coupling matrix $K$ for $N=128$ and $J^*=0.2$. \ref{sub@fig:AntipodalDefect_NN_Dif}) A plot of the nearest neighbor interactions for $K_A^1-K_A^{0.2}$, where $K_A^1$ is the $K_A$ matrix for no defect. The x axis is the lattice distance from the defect ($m=0$). The impact of the defect on the nearest neighbor hoppings is highly local.}\label{fig:AntipodalDefectPlotCluster1}
	\end{figure}

	We begin by looking at the case where we have two energy defects of equal strength on opposite sides of our periodic chain (hence, antipodal). Our subsystem $A$ is half of the chain with one of the defects at its center. This is shown in figure \ref{fig:AntipodalDefectSetup}. Calculating $K_A$ for a series of defect strengths and system sizes (see figure \ref{fig:AntipodalDefectPlotCluster1}), we find that the entanglement Hamiltonian is dominated by nearest neighbor interactions, whose distribution largely resembles that of the no defect case. The exception to this behavior takes place around the defect itself, where the bonds have deformed.  This effect is highly localized around the defect, and after only a few lattice sites the bonds have returned to their no defect values. 

\begin{figure}[h]
		\centering
		\includegraphics[scale = 0.7]{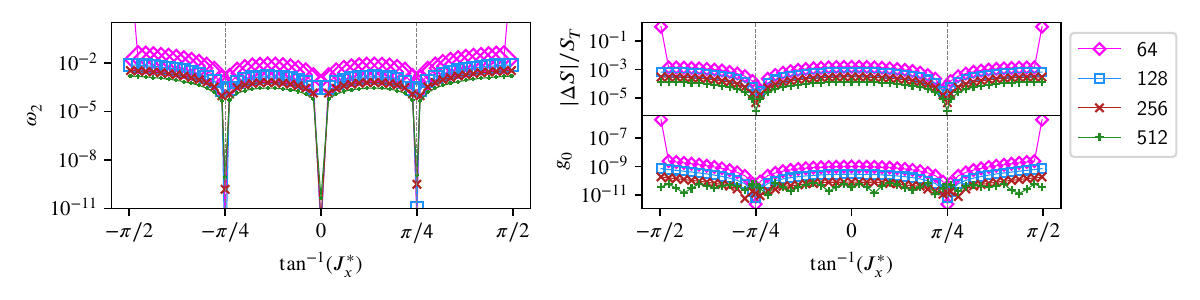}
		\caption{(Left) The weights of the second operator $L_2$ for the antipodal defects. Note that these values, after rescaling by $4L$ are relatively system size independent, whereas $\omega_1$ is linear in $L$. (Right Top) The relative difference in entanglement entropy. This agreement is best for $J^* = \pm 1$ and improves as the system size grows. (Right Bottom) The energy fluctuation $g_0$ for the antipodal defects. We see in all plots that for the large defect, $J^*\approx \pm 100$, this reconstruction procedure has failed for $N=64$.} \label{fig:APDefectsFitFig}
	\end{figure}

	If we perform entanglement Hamiltonian reconstruction with the operators $L_1$ and $L_2$, we find that the $\mathcal{K}_A$ is well approximated by 
	\begin{equation}
		\tilde{\mathcal{K}}_A = 4L\sum_{n=0}^{2L-2}\sin\brac{\frac{\pi(n+1)}{2L}}\gamma_n\gamma_{n+1}
	\end{equation}	
	 as $\omega_2<<\omega_1$ and decreases as $N$ grows (figure \ref{fig:APDefectsFitFig}). We also see that $g_0$ is $\mathcal{O}(10^{-9})$ or smaller for large $N$, with minimal variation with system size. We also see that the relative difference in the entanglement entropy decreases with system size, showing that our reconstructed $\tilde{K}_A$ and the actual $K_A$ agree. If we look at the infinite system results from \cite{Mintchev_Tonni_2021}, we see that the form of the local terms make sense, as this is just the Hamiltonian weighted by a conformal factor. The lack of non-local interactions is also reasonable, as while there is still communication through $\bar{A}$, this communication is interrupted by the defect. This disrupted communication is also why the negativity for the antipodal systems is less than the single defect in figure \ref{fig:NegativityPlot}. If we remove this defect, it will be reasonable to expect persistent non-local interactions.

\section{Single Centered Defect}\label{sec:SingleCenteredDefect}	
	\begin{figure}[H]
		\centering
		\begin{tikzpicture}[scale = 0.65]
			\tikzmath{
			coordinate \c;
			coordinate \a;
			coordinate \b;
			\AngleStep = 22.5;
			\radius = 2.5;
			\spinsize = 0.1;
			\majoranasep = 5;
			\majoranasize = 0.1;
			\JOffsetSpins = 1.0;
			\JOffsetMajoranas = 1.0;
			\gWidth = 2;
			\JWidth = 2;
			\sigmazoffset = 4;
			\verticalshift = -2-\radius;
			\a = (\AngleStep-\majoranasep/2-\majoranasize*1.41cm-\AngleStep/4: {\JOffsetMajoranas*\radius});
			\b = (\AngleStep-\majoranasep/2+180-\majoranasize*1.41cm-\AngleStep/4: {\JOffsetMajoranas*\radius+0.5});
			{\draw[line width = 6, color = black!40] (\a) arc[start angle = \AngleStep-\majoranasep/2-\majoranasize*1.41cm-\AngleStep/4, end angle = \AngleStep-\majoranasep/2+180-\majoranasize*1.41cm-\AngleStep/4, radius = \radius];};
			for \x in {\AngleStep,2*\AngleStep,...,180} {
				\a = (\x+\majoranasep: {\JOffsetMajoranas*\radius});
				\b = (\x-\majoranasep+\AngleStep: {\JOffsetMajoranas*\radius});
				{\draw[line width=\JWidth, JBond] (\a) -- (\b);};
				\a = (\x+\majoranasep+180: {\JOffsetMajoranas*\radius});
				\b = (\x-\majoranasep+\AngleStep+180: {\JOffsetMajoranas*\radius});
				{\draw[line width=\JWidth, JBond] (\a) -- (\b);};
			};
			\a = (\majoranasep+180: {\JOffsetMajoranas*\radius});
			\b = (0-\majoranasep+\AngleStep+180: {\JOffsetMajoranas*\radius});
			for \x in {0,\AngleStep,...,360} {
				\a = (\x-\majoranasep: {\JOffsetMajoranas*\radius});
				\b = (\x+\majoranasep: {\JOffsetMajoranas*\radius});
				{\draw[line width=\gWidth, gBond] (\a) -- (\b);};
			};
			for \x in {0,\AngleStep,...,360} {
				\c = (\x+\majoranasep: {\radius});
				{ \fill (\c) circle [radius={\majoranasize} ]; };
				\c = (\x-\majoranasep: {\radius});
				{\node[rectangle, draw, rotate = \x-\majoranasep, fill = black, minimum width=\majoranasize*1.41cm,minimum height=\majoranasize*1.41cm, inner sep=2pt] at (\c) {};};
			};
			\a = (\AngleStep-\majoranasep/2-\majoranasize*1.41cm-\AngleStep/4: {\JOffsetMajoranas*\radius+0.5});
			\b = (\AngleStep-\majoranasep/2+180-\majoranasize*1.41cm-\AngleStep/4: {\JOffsetMajoranas*\radius+0.5});
			\a = (90+\AngleStep-\majoranasep/2-\majoranasize*1.41cm-\AngleStep/4: {\JOffsetMajoranas*\radius+0.5});
			\b = (90+\AngleStep-\majoranasep/2-\majoranasize*1.41cm-\AngleStep/4: {\JOffsetMajoranas*\radius-0.5});
			{\draw[thick, decoration = zigzag, decorate] (\a) -- (\b);};
			{\node[below] at (\b) {$J^*$, $x=\tfrac{1}{2}$};};
			\a = (45+\AngleStep-\majoranasep/2-\majoranasize*1.41cm-\AngleStep/4: {\JOffsetMajoranas*\radius+1});
			{\node[below] at (\a) {$A$};};
			}
		\end{tikzpicture}
		\caption{Diagram for the setup for the defect in $A$.}\label{fig:SingleDefectSetup}
	\end{figure}
	\begin{figure}
		\centering\hspace{-4cm}
		\begin{minipage}[c]{0.8\textwidth}
			\subcaptionbox{\label{fig:SingleDefect_NN_Ns}}{\includegraphics[scale = 0.7]{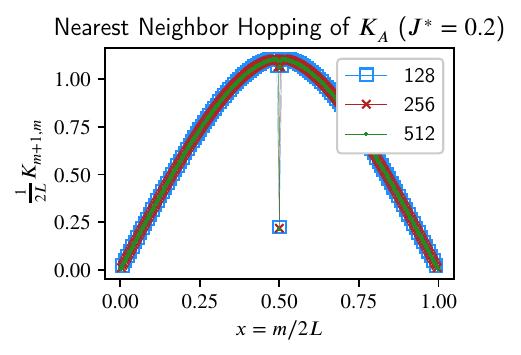}}
			\subcaptionbox{\label{fig:SingleDefect_NN_Defects}}{\includegraphics[scale = 0.7]{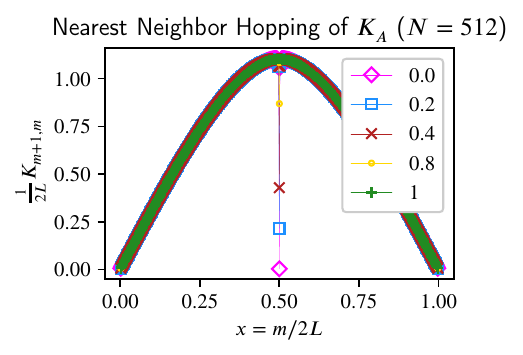}}
		\end{minipage}
		\begin{minipage}[c]{0.1\textwidth}
			\subcaptionbox{\label{fig:SingleDefect_K}}{\includegraphics[scale = 0.7]{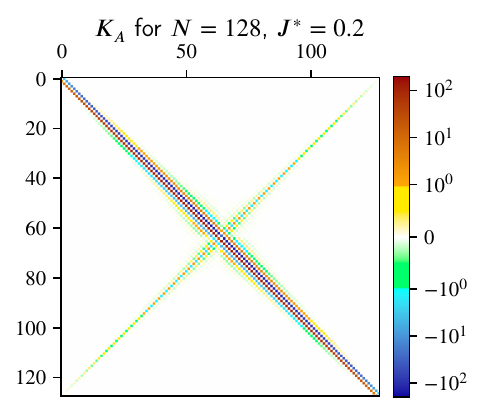}}
		\end{minipage}
		\begin{minipage}[c]{0.45\textwidth}
			\subcaptionbox{\label{fig:SingleDefect_NN_Dif}}{\includegraphics[scale = 0.7]{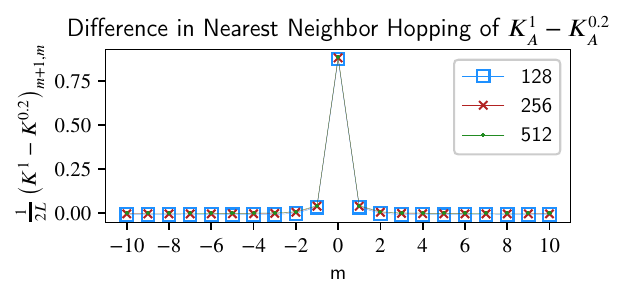}}
		\end{minipage}
		\caption{\ref{sub@fig:SingleDefect_NN_Ns}) Plot of nearest neighbor interactions for various system sizes for $J^* = 0.2$. \ref{sub@fig:SingleDefect_NN_Defects}) Plot of nearest neighbor interactions for various defects for $N=512$. \ref{sub@fig:SingleDefect_K}) The full coupling matrix $K$ for $N=128$ and $J^*=0.2$. The scale is symmetric log where the linear threshold is $\pm1$. Note the prominent anti-diagonal entries, which are symmetric hopping terms across the defect. \ref{sub@fig:SingleDefect_NN_Dif}) A plot of the nearest neighbor interactions for $K^1-K^{0.2}$, where $K^1$ is the $K$ matrix for no defect. The x axis is the lattice distance from the defect ($m=0$). Note that regardless of system size, the impact of the defect on the nearest neighbor interactions is only a couple lattice sites. }\label{fig:SingleDefectPlotCluster1}
	\end{figure}
\begin{figure}
		\centering
		\begin{minipage}[c]{0.45\textwidth}
			\subcaptionbox{\label{fig:SingleDefectSymmtricHopping_Ns}}{\includegraphics[scale = 0.7]{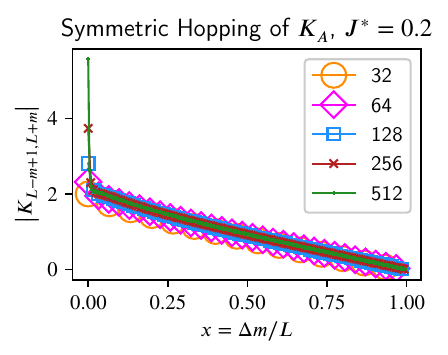}}
		\end{minipage}
		\begin{minipage}[c]{0.45\textwidth}
			\subcaptionbox{\label{fig:SingleDefectSymmtricHopping_Defects}}{\includegraphics[scale = 0.7]{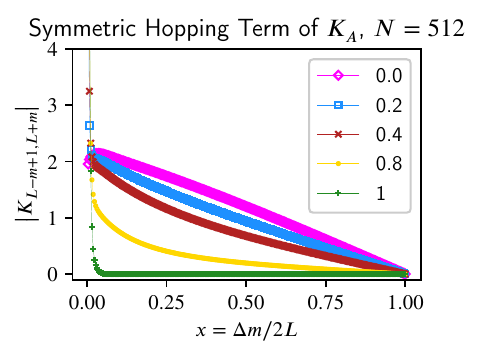}}
		\end{minipage}
		\caption{Plots of the symmetric hopping term $K_{L-m+1,L+m}$. Note that in these plots, the defect term itself is left out ($m=0$). \ref{sub@fig:SingleDefectSymmtricHopping_Ns}) Symmetric hopping term for $J^*=0.2$ for various system sizes. After a few lattice sites away from the defect, the magnitude of these terms does not depend on the system size (there is no factor of $1/2L$ rescaling $K$, unlike in plots of the nearest neighbor terms). \ref{sub@fig:SingleDefectSymmtricHopping_Defects}) Plots of the symmetric hopping terms for various defect strengths. These terms grow as $J^*\to0$. It is also important to note that for both of these plots, we are looking at $\left|K_{L-m+1,L+m}\right|$ and there is actually an alternating minus sign.  }\label{fig:SingleDefectSymmetricHopping}
	\end{figure}

	Removing the defect in $\bar{A}$, we obtain the geometry shown in figure \ref{fig:SingleDefectSetup}. Calculating the lattice entanglement Hamiltonians, figure \ref{fig:SingleDefectPlotCluster1}, we see that we have local terms that look identical to the antipodal defects, but the plot of $K_A$ shows stark non-local interactions. Focusing in on this (figure \ref{fig:SingleDefectSymmetricHopping}), we see that these non-local terms are relatively system size independent, especially as their distance grows, and that the magnitude and shape of these terms has a non-trivial dependence on the defect strength.

	\begin{figure}
		\centering
		\includegraphics[scale = 0.7]{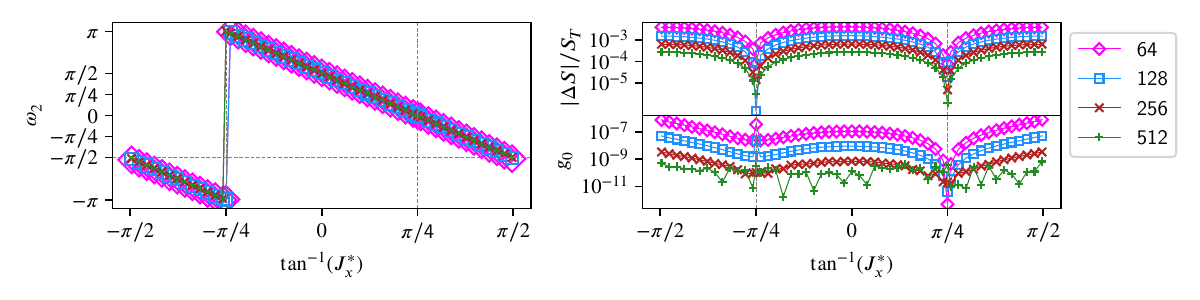}
		\caption{(Left) The weights $\omega_2(b)$ of the operator $L_2$. These values are system size follow the equation given in \ref{eq:SingleDefectOmega2}. (Right Top) The relative difference in entanglement entropy. (Right Bottom) The energy fluctuation $g_0$ for the single defect. For large systems, both of these show high agreement between the actual lattice $K_A$ and the predicted CFT result. We also point out that these results include the antiperiodic defect, $J^*=-1$, which introduces a zero mode into the system as well as cause a degeneracy in the entanglement spectrum. This introduces some numerical issues, hence the rise in $g_0$ at $\tan^{-1}(J^*) = -\pi/4$}
	\end{figure}

\begin{figure}
		\centering
		\begin{minipage}[c]{\paperwidth}
			\hspace{0cm}
			\subcaptionbox{\label{fig:PlotOfKAntiperiodicHalfDefect}}{\includegraphics[scale = 0.6]{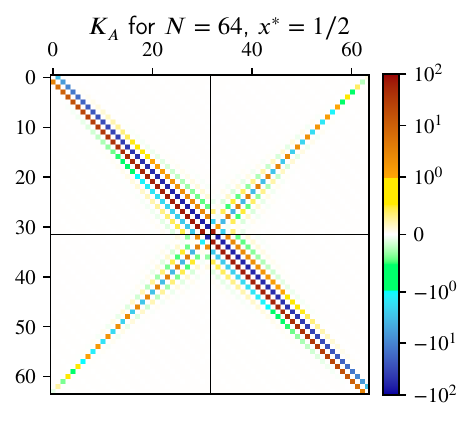}}
			\subcaptionbox{\label{fig:PlotOfKAntiperiodicOneQuarterDefect}}{\includegraphics[scale = 0.6]{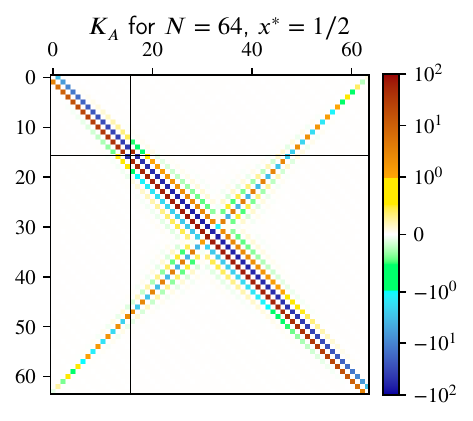}}
			\subcaptionbox{\label{fig:PlotOfKAntiperiodicOutsideDefect}}{\includegraphics[scale = 0.6]{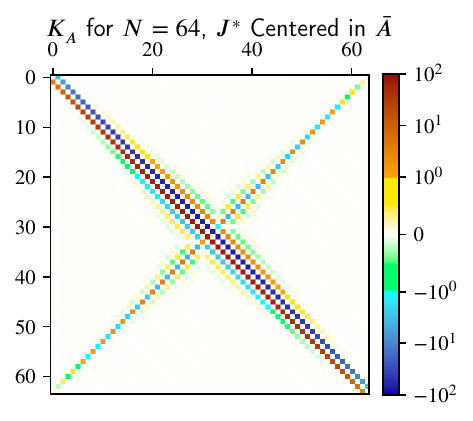}}
		\end{minipage}
		\caption{Plots of antiperiodic systems for different defect locations: Centered in $A$ (\ref{sub@fig:PlotOfKAntiperiodicHalfDefect}), off-centered in $A$ (\ref{sub@fig:PlotOfKAntiperiodicOneQuarterDefect}) and centered in $\bar{A}$ (\ref{sub@fig:PlotOfKAntiperiodicOutsideDefect}). Note that for when the defect is in $A$, all the bonds across the defect pick up a relative minus sign compared to when the defect is outside the subsystem. It is also worth noting that, while for \ref{sub@fig:PlotOfKAntiperiodicOutsideDefect} the defect is centered in $\bar{A}$, $K$ is the same regardless of where the defect is, as long as it is not a bond within $A$. The symmetric hopping term (up to the minus sign acquired if coupling across the defect) is the same.}\label{fig:AntiperiodicForDifferentPositions}
	\end{figure}

	If we perform the Hamiltonian reconstruction, we find that our choice of operators yields very good fits, with $g_0$ being $\mathcal{O}(10^{-9})$ for our largest system sizes. The most interesting part of this fit is the dependence of $\omega_2$ on the defect strength. If we define $b=\tan^{-1}(J^*)$, then we see that $\omega_2$ goes as 
	\begin{equation}\label{eq:SingleDefectOmega2}
		\omega_2(b) = -2(b+\sfrac{\pi}{4}\mod\pi)+\pi,
	\end{equation}
	or 
	\begin{equation}
		J^* = -\tan\brac{\frac{\omega_2-\pi}{2}+\frac{\pi}{4}}.
	\end{equation}
	With the single defect, we can also ask about the behavior of $K_{\bar{A}}$. The resulting entanglement Hamiltonian is very similar to $K_{A}$ except without the dip around the defect and some relative minus signs along the anti-diagonals. If we perform entanglement Hamiltonian construction of $K_{\bar{A}}$, we find that the fit and weights behave the same as $K_A$ up to an overall minus sign in $\omega_2(b)$. From this, we propose that the continuum limit entanglement Hamiltonian for a single defect centered in $A$ is given by
	\begin{equation}\label{eq:Omega2InandOut}
		\begin{aligned}
			\mathcal{K}^{\text{CFT}}_A &= 4L L_1 +\brac{\pi-2(b+\sfrac{\pi}{4}\mod\pi)}L_2,\\
			\mathcal{K}^{\text{CFT}}_{\bar{A}} &= 4L L_1 -\brac{\pi-2(b+\sfrac{\pi}{4}\mod\pi)}L_2.
		\end{aligned}
	\end{equation}
	For the antiperiodic system ($J_1^*=-1$), we also see that symmetric hopping term persists, regardless of where the defect is located (figure \ref{fig:AntiperiodicForDifferentPositions}). The only difference is a factor of $-1$ depending on whether or not a bond crosses the defect. We can modify our definition of $L_2$ to include this factor:
	\begin{equation}
		\tilde{L}_2 = \sum_{n=0}^{L-1}(-1)^{n+1-\Theta(2 X+1-n)}\sin\brac{\frac{\pi(n+1)}{2L}}\gamma_n\gamma_{2L-n-1},
	\end{equation}
	where $X\in [0,N]$ is the location of the defect in terms of the $J$ couplings (with $X=0$ being the first bond in $A$ and $N-1$ being the bond going into $A$) and $\Theta(x)$ is the Heaviside Theta function. In this form, any bond that crosses the defect picks up a factor of $-1$. This form also, for $J^*=-1$, reproduces the extra factor of $-1$ in $\omega_2$ for $K_{\bar{A}}$, as none of the bonds cross the defect (versus all for $K_A$), resulting in an overall factor of $-1$ in $\omega_2$. This only works for the antiperiodic case, as when $J^* \ne -1$, off-center defects introduce more complicated terms. This is the topic of \ref{sec:SingleOffCenteredDefect}.

\begin{figure}[h]
		\centering
		\includegraphics[scale = 0.7]{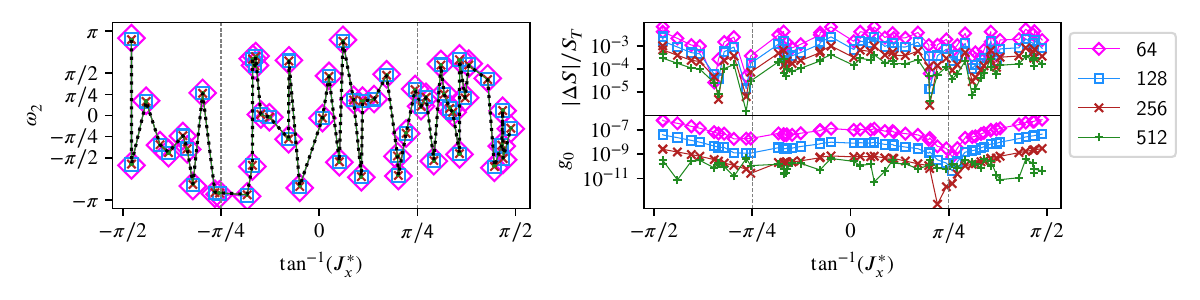}
		\caption{The weights and agreement for random choices of $J_1^*$ and $J_2^*$, sorted by $J_1^*$. The black line in plot of $\omega_2$ is the predictions from equation \ref{eq:GeneralizedOmega2}.} \label{fig:UAPDefectsFitFig}
	\end{figure}

\begin{figure}[h]
		\centering
		\includegraphics[scale = 0.7]{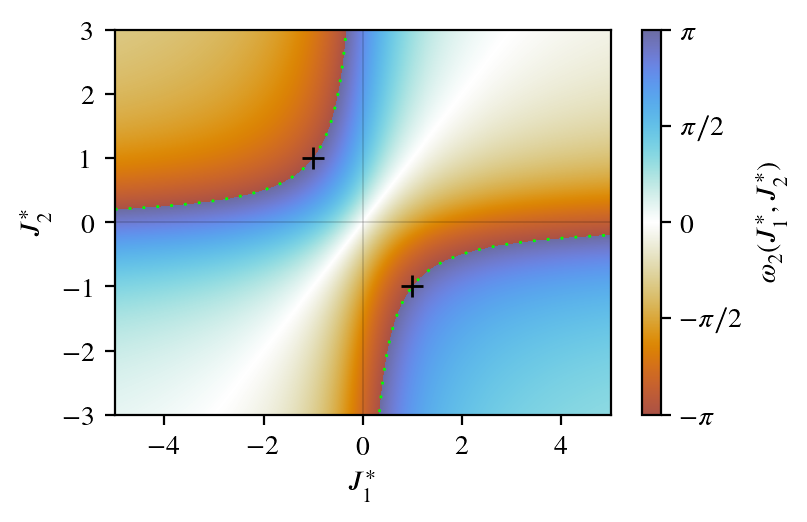}
		\caption{Plot of $\omega_2(J_1^*, J_2^*)$. There is a discontinuity at $J_2^* = -1/J_1^*$ (green dotted line) where $\omega_2$ goes from $-\pi$ to $\pi$. This is also the point where $|\omega_2(J_1^*, J_2^*)|$ is maximal; meaning that along this line, the entanglement Hamiltonian is the most non-local. This is not unexpected as all of these models contain zero modes, with the canonical antiperiodic system lying at $(1,-1)$ and $(-1,1)$, denoted by the ``$+$'' markers.} \label{fig:omegas}
	\end{figure}

	We can generalize \ref{eq:SingleDefectOmega2} to the case where we have two centered antipodal but not necessarily equal defects and recover the results from \ref{sec:AntipodalDefect} as well as the behavior of $\omega_2$ for $K_A$ vs $K_{\bar{A}}$ from equation \ref{eq:Omega2InandOut}. After performing entanglement Hamiltonian reconstruction for a variety of $J_1^*$ and $J_2^*$, we find that $\omega_2$ follows the form
	\begin{equation}\label{eq:GeneralizedOmega2}
		\omega_2(b_1, b_2) = -2\brac{\para{(b_1+\sfrac{\pi}{4})- (b_2-\sfrac{\pi}{4})}\mod\pi}+\pi
	\end{equation}
	where $b_i = \tan^{-1}(J_i^*)$. The results for this can be seen in figure \ref{fig:UAPDefectsFitFig}. In this form, the single defect in $A$ is given by $J_2^*=1\implies b_2 = \pi/4$ and we recover equation \ref{eq:SingleDefectOmega2}, while $J_1^*=J_2^*$ results in $\omega_2(b,b) = 0$, recovering the results of section \ref{sec:AntipodalDefect}. Also, if our single defect is in $\bar{A}$, it can be shown that $\omega_2(1,b) = -\omega_2(b,1)$, which is the factor of $-1$ seen between the non-local terms of  $K_A$ and $K_{\bar{A}}$. The last interesting configuration of defects is when $J_2^* = -1/J_1^*$, resulting in $\omega_2 = \pm\pi$, depending on whether $J_1^*$ is positive ($+\pi$) or negative ($-\pi$). This can be seen in the plot of $\omega_2(J_1^*, J_2^*)$ in figure \ref{fig:omegas}.  

	On a last note, for this section we have chosen to define $b_i = \tan^{-1}(J_i^*)$, however we could have chosen $\tilde{b}_i = \cot^{-1}(J_i^*)$, which is the parameterization used in \cite{Roy2022,RogersonPollmannRoy2022,BrehmBrunner2015,Peschel_2012} when studying single boundary defects. With this definition, we get
	\begin{equation}
		\omega_2(\tilde{b}_1, \tilde{b}_2) = 2\brac{\para{(\tilde{b}_1+\sfrac{\pi}{4})- (\tilde{b}_2-\sfrac{\pi}{4})}\mod\pi}-\pi.
	\end{equation}
	We chose $\tan^{-1}$ as it puts $J^* = \pm 1$ at $b = \pm\pi/4$ and $0$ at $0$, which is more intuitive when plotting. 
\section{Conclusion}
In this work we have explored the lattice entanglement Hamiltonian for a variety of single and antipodal defect cases. In the main text, we have focused on systems where we have been able to provide an ansatz for the form of the continuum limit entanglement Hamiltonian. From these, we have been able to characterize the behavior of the non-local terms as a function of the defect strengths. In the appendix, we have looked at configurations where we are unable to provide an ansatz for their continuum limit couplings. As a next step, we would like to employ a more comprehensive Hamiltonian reconstruction technique that will allow us to fit a large set of individual couplings (as done in \cite{PhysRevB.99.235109} for no defect). This was attempted, even for the centered defects, however adding in the necessary long range couplings leads to highly degenerate solutions and we were unable to make meaningful predictions. We believe that a method that utilizes higher entanglement modes will allow us to enforce additional constraints to deal with this degeneracy issue, giving us the ability to study more complicated defect configurations.

	\section*{Acknowledgments}
	The author would like to thank my advisor Professor Ananda Roy and my colleague David Rogerson for their great and insightful discussions. 
\appendix

\section{A Fermionic Logarithmic Negativity Argument for Non-Local Interactions}\label{sec:FLNArgument}
	In sections \ref{sec:AntipodalDefect} and \ref{sec:SingleCenteredDefect} we will use the fermionic logarithmic negativity (FLN)\cite{FermionicNegativity_2017,FermionicNegativity_2019} to make an argument about the presence of non-local interactions across a defect. In this subsection, we will outline the FLN, describe how to calculate it using free fermions and argue why a non-zero FLN across a defect implies these extra couplings. Beginning with the normal logarithmic negativity, if we have a system with two subsystems $A$ and $B$, along with a state $\ket{\psi}$, the entanglement between $A$ and $B$ can usually be captured by the quantity 
	\begin{equation}
		\mathcal{E}(\rho_{AB}) = \log\abs{\abs{\rho_{AB}^{T_A}}}.
	\end{equation}
	where $\rho_{AB}^{T_A}$ is the partial transpose of the reduced density matrix with respect to the subsystem $A$. If $\mathcal{E}$ is non-zero, there is entanglement between $A$ and $B$ for the given state.\footnote{Note that $\mathcal{E}\ne0$ implies entanglement, however $\mathcal{E}=0$ does not imply a lack of entanglement in general.} The normal partial transpose does not take into account exchange statistics for fermionic systems, which is where the fermionic partial transpose comes in. It takes into account the fermionic exchange statistics and allows us to define a fermionic version of the logarithmic negativity. 

	The next few sections are a summary of the derivation of the FLN following \cite{HowIGetFLN}, the implications of the FLN come after the formula in equation \ref{eq:FLN}. If we have a subsystem $A = A_1\cup A_2$ and we are partially transposing with respect to $A_1$, we can write the restricted covariance matrix as 
	\begin{equation}
		\Gamma_A = \begin{pmatrix} \Gamma_{1} & \Gamma_{12} \\ \Gamma_{21} & \Gamma_2\end{pmatrix}
	\end{equation}
	where $\Gamma_1,\Gamma_2$ are the restricted covariance matrices to $A_1$ and $A_2$ respectively. $\Gamma_{12} = \Gamma_{21}^\dagger$ is the interactions between the two subsystems. We can now define 
	\begin{equation}
		\Gamma_A' = \begin{pmatrix} -\Gamma_{1} & \i\Gamma_{12} \\ \i\Gamma_{21} & \Gamma_2\end{pmatrix}
	\end{equation}
	where $\Gamma_A'$ is the covariance matrix associated with the partially transposed state. From there, define $U = (-\mathbb{I}_1)\oplus\mathbb{I}_2$ and the covariance matrix associated with the fermionic partial transpose is given by 
	\begin{equation}
		\tilde{\Gamma}_A = \tanh\brac{\frac{1}{2}\log\para{\frac{\mathbb{I} - \Gamma_A'}{\mathbb{I} + \Gamma_A'}U}}.
	\end{equation}
	The eigenvalues of $\tilde{\Gamma}_A$ come in pairs, $\pm\tilde\nu_i$ and we define the fermionic logarithmic negativity as
	\begin{equation}\label{eq:FLN}
		\mathcal{N}=\sum_{\curly{\tilde\nu_i}}\log\brac{\abs{\frac{1-\tilde\nu_i}{2}}+\abs{\frac{1+\tilde\nu_i}{2}}}+\frac{1}{2}\log{\det\Gamma_1}.
	\end{equation}
	\begin{figure}[h]
		\centering
		\includegraphics[scale = 0.7]{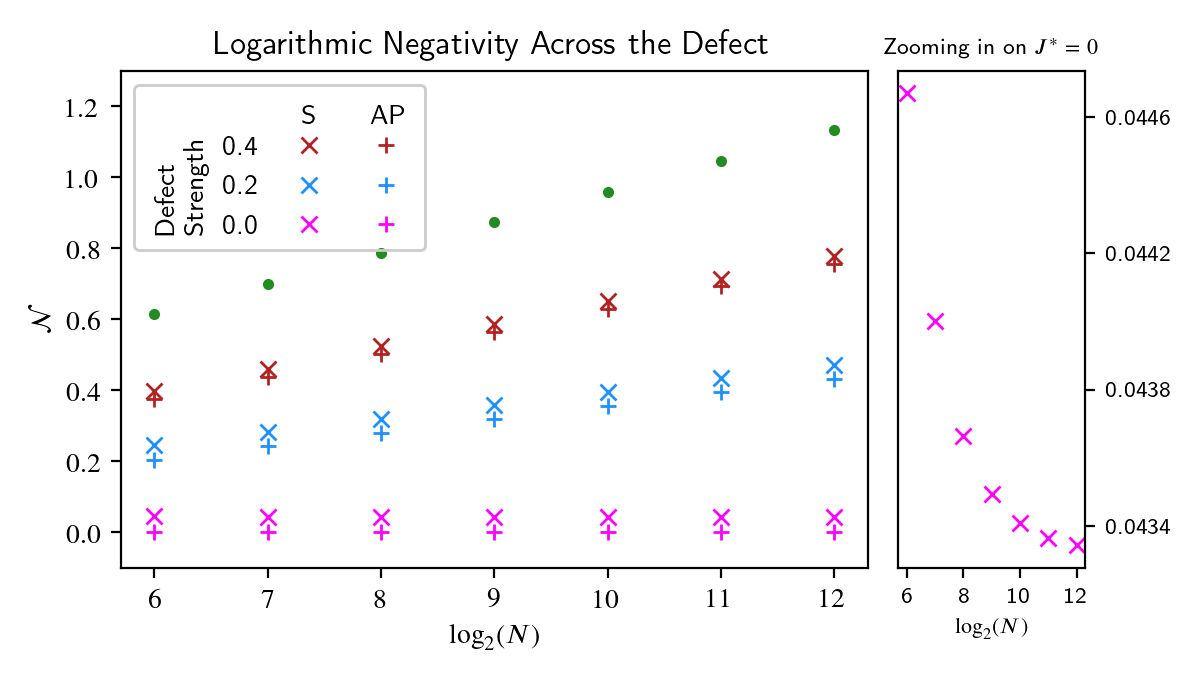}
		\caption{The fermionic logarithmic negativity across the defect for the single centered defect (S) and the antipodal centered defects (AP) along with the case with no defect (green circles). As the couplings decrease, the gap in the entanglement entropy grows and for $J^*=0$, the negativity for the antipodal case is numerically zero. For the single defect however, the negativity is asymptotically finite, though small.}\label{fig:NegativityPlot}
	\end{figure}
	Notably, if $\Gamma_A = \Gamma_1\oplus\Gamma_2$, then $\mathcal{N} = 0$. In terms of $K_A$, this implies that if $K_A = K_{A_1}\oplus K_{A_2}$, then the negativity is zero. We can rephrase this as: if $\mathcal{N}\ne 0$, then $\Gamma_A \ne \Gamma_1\oplus\Gamma_2$. In other words, if we have non-zero fermionic logarithmic  negativity, then we must have inter-subsystem couplings in both $\Gamma_A$ and $K_A$. In general, this coupling is generated by the bond connecting $A_1$ and $A_2$. However, if this coupling is zero, then the power series of the local CFT terms (from equation \ref{eq:K_Lat}) will not have any cross-coupling terms. These terms must then be generated by communication through $\bar{A}$, and we see from figure \ref{fig:NegativityPlot} that this results in asymptotically non-zero negativity. This means that for $J^*=0$, there must be non-local terms in $K_A$ that persist in the scaling limit. We can hypothesize that these terms will have a different scaling relation than the local terms, as the negativity generated by a non-zero defect grows logarithmically with system size, whereas the $J^*=0$ negativity is, at first order, constant with system size. This ends up being the case, as the local interactions scale linearly in system size whereas the non-local interactions are system size independent.   

\section{Single Off-Centered Defect}\label{sec:SingleOffCenteredDefect}
\begin{figure}[h]
		\centering\hspace{-5cm}
		\begin{minipage}[c]{0.8\textwidth}
			\subcaptionbox{\label{fig:OffCenterDefectNN}}{\includegraphics[scale = 0.7]{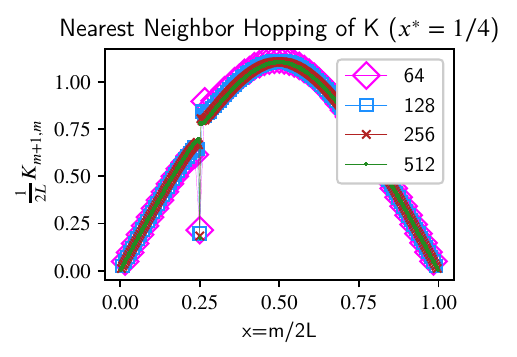}}
			\subcaptionbox{\label{fig:OffCenterDefect_xs}}{\includegraphics[scale = 0.7]{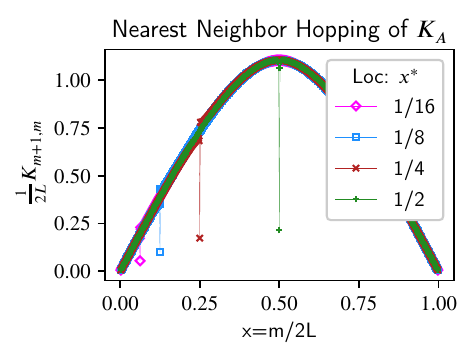}}
		\end{minipage}
		\begin{minipage}[c]{0.1\textwidth}
			\subcaptionbox{\label{fig:OffCenterDefect_K}}{\includegraphics[scale = 0.7]{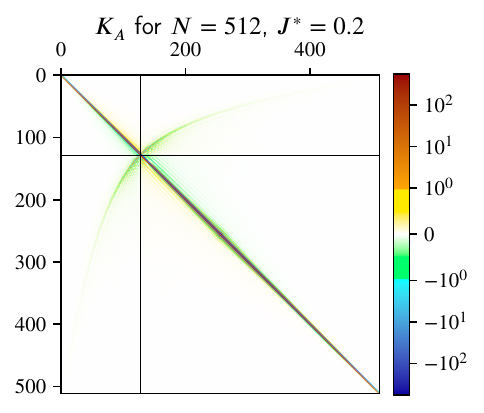}}
		\end{minipage}
		\caption{\ref{sub@fig:OffCenterDefectNN}) Plot of the nearest neighbor hopping terms of $K_A$ for multiple system sizes with the defect, $J^* = 0.2$,  at $x=1/4$. \ref{sub@fig:OffCenterDefect_xs}) Plot of the nearest neighbor hopping terms for multiple defect locations ($N=512$, $J^* = 0.2$). \ref{sub@fig:OffCenterDefect_K}) Symmetric log plot of $K_A$. The black lines indicate the defect bond.}\label{fig:SomePlotsForOffCenterDefect1}
	\end{figure}

	If we consider a single defect off-center in $A$, we find that the local couplings only are affected within a few lattice sites of the defect (figures \ref{fig:OffCenterDefectNN},\ref{sub@fig:OffCenterDefect_xs}). In figure \ref{fig:OffCenterDefect_K}, we see that the cross defect coupling terms are non-trivial. Varying the position of this defect yields a complicated dependence on the position of the defect. 

\begin{figure}[h]
		\centering
		\includegraphics[scale = 0.7]{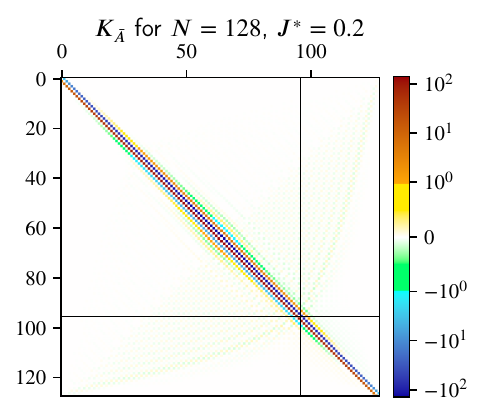}
		\caption{$K_{\bar{A}}$ for the off-centered defect. The subsystem is the interval $[N/2, N-1]$.}\label{fig:OffCenter_Outside_K}
	\end{figure}

	If we look at $K_{\bar{A}}$ (figure \ref{fig:OffCenter_Outside_K}), we see that the non-local defects are weaker than for $K_A$ but are on the side closest to the defect. If we were to look at the antipodal case with off-center defects, we find that $K^{\text{(AP)}}_A$ looks like a combination of $K^{\text{(S)}}_A$ and $K^{\text{(S)}}_{\bar{A}}$, with the stronger non-local terms being the ones around the defect in $A$. 
\section{Single Boundary Defect}\label{sec:SingleBoundaryDefect}
	\begin{figure}[h]
		\centering
		\begin{tikzpicture}[scale = 0.65]
			\tikzmath{
			coordinate \c;
			coordinate \a;
			coordinate \b;
			\AngleStep = 22.5;
			\radius = 2.5;
			\spinsize = 0.1;
			\majoranasep = 5;
			\majoranasize = 0.1;
			\JOffsetSpins = 1.0;
			\JOffsetMajoranas = 1.0;
			\gWidth = 2;
			\JWidth = 2;
			\sigmazoffset = 4;
			\verticalshift = -2-\radius;
			\a = (\AngleStep-\majoranasep/2-\majoranasize*1.41cm-\AngleStep/4: {\JOffsetMajoranas*\radius});
			\b = (\AngleStep-\majoranasep/2+180-\majoranasize*1.41cm-\AngleStep/4: {\JOffsetMajoranas*\radius+0.5});
			{\draw[line width = 6, color = black!40] (\a) arc[start angle = \AngleStep-\majoranasep/2-\majoranasize*1.41cm-\AngleStep/4, end angle = \AngleStep-\majoranasep/2+180-\majoranasize*1.41cm-\AngleStep/4, radius = \radius];};
			for \x in {\AngleStep,2*\AngleStep,...,180} {
				\a = (\x+\majoranasep: {\JOffsetMajoranas*\radius});
				\b = (\x-\majoranasep+\AngleStep: {\JOffsetMajoranas*\radius});
				{\draw[line width=\JWidth, JBond] (\a) -- (\b);};
				\a = (\x+\majoranasep+180: {\JOffsetMajoranas*\radius});
				\b = (\x-\majoranasep+\AngleStep+180: {\JOffsetMajoranas*\radius});
				{\draw[line width=\JWidth, JBond] (\a) -- (\b);};
			};
			\a = (\majoranasep+180: {\JOffsetMajoranas*\radius});
			\b = (0-\majoranasep+\AngleStep+180: {\JOffsetMajoranas*\radius});
			for \x in {0,\AngleStep,...,360} {
				\a = (\x-\majoranasep: {\JOffsetMajoranas*\radius});
				\b = (\x+\majoranasep: {\JOffsetMajoranas*\radius});
				{\draw[line width=\gWidth, gBond] (\a) -- (\b);};
			};
			for \x in {0,\AngleStep,...,360} {
				\c = (\x+\majoranasep: {\radius});
				{ \fill (\c) circle [radius={\majoranasize} ]; };
				\c = (\x-\majoranasep: {\radius});
				{\node[rectangle, draw, rotate = \x-\majoranasep, fill = black, minimum width=\majoranasize*1.41cm,minimum height=\majoranasize*1.41cm, inner sep=2pt] at (\c) {};};
			};
			\a = (\AngleStep-\majoranasep/2-\majoranasize*1.41cm-\AngleStep/4: {\JOffsetMajoranas*\radius+0.5});
			\b = (\AngleStep-\majoranasep/2+180-\majoranasize*1.41cm-\AngleStep/4: {\JOffsetMajoranas*\radius+0.5});
			\a = (180+\AngleStep-\majoranasep/2-\majoranasize*1.41cm-\AngleStep/4: {\JOffsetMajoranas*\radius+0.5});
			\b = (180+\AngleStep-\majoranasep/2-\majoranasize*1.41cm-\AngleStep/4: {\JOffsetMajoranas*\radius-0.5});
			{\draw[thick, decoration = zigzag, decorate] (\a) -- (\b);};
			{\node[right] at (\b) {$J^*$};};
			\a = (45+\AngleStep-\majoranasep/2-\majoranasize*1.41cm-\AngleStep/4: {\JOffsetMajoranas*\radius+1});
			{\node[below] at (\a) {$A$};};
			}
		\end{tikzpicture}
		\caption{Diagram for the boundary defects.}\label{fig:BoundaryDefectSetup}
	\end{figure}
	\FloatBarrier
	\begin{figure}[h]
		\centering
		\begin{minipage}[c]{0.45\textwidth}
			\subcaptionbox{\label{fig:SingleDefectOnBoundary_NN_Ns}}{\includegraphics[scale = 0.7]{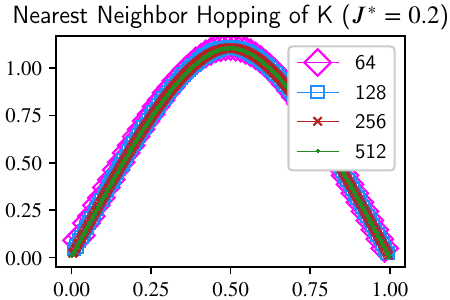}}
		\end{minipage}
		\begin{minipage}[c]{0.45\textwidth}
			\subcaptionbox{\label{fig:SingleDefectOnBoundary_K}}{\includegraphics[scale = 0.7]{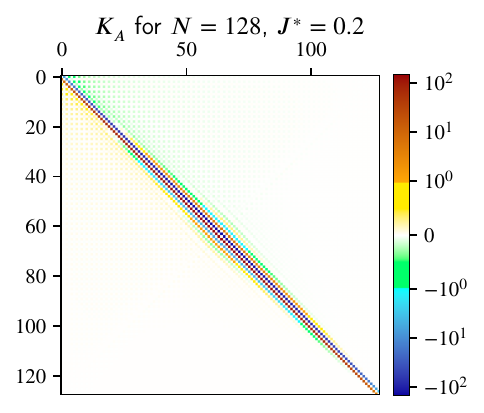}}
		\end{minipage}
		\caption{\ref{sub@fig:SingleDefectOnBoundary_NN_Ns}) Plot of nearest neighbor couplings for the defect on the subsystem boundary. Note that the effect on the nearest neighbor hoppings is minimal. \ref{sub@fig:SingleDefectOnBoundary_K}) Plot of $K$ for $J^*=0.2$ as a comparison. Unlike previous Ks, this matrix is highly non-local around the boundary with the defect. }\label{fig:SingleDefectOnBoundaryCluster1}
	\end{figure}
\begin{figure}[h]
		\centering
		\begin{minipage}[c]{\paperwidth}
			\hspace{-2cm}
			\subcaptionbox{\label{fig:PlotofKSingleEdgeDefect_N64_-5}}{\includegraphics[scale = 0.6]{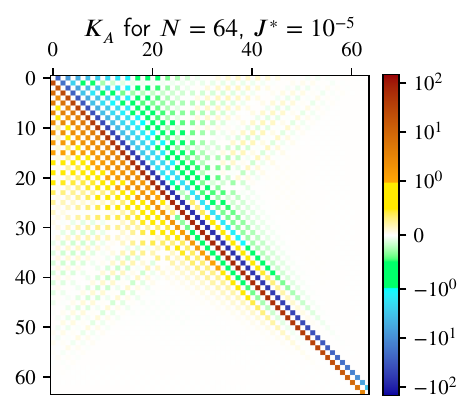}}
			\subcaptionbox{\label{fig:PlotofKSingleEdgeDefect_N64_-10}}{\includegraphics[scale = 0.6]{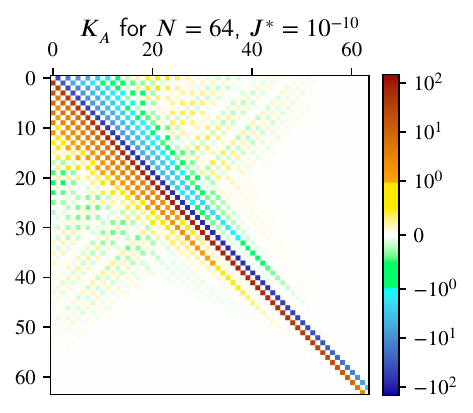}}
			\subcaptionbox{\label{fig:PlotofKSingleEdgeDefect_N64_-50}}{\includegraphics[scale = 0.6]{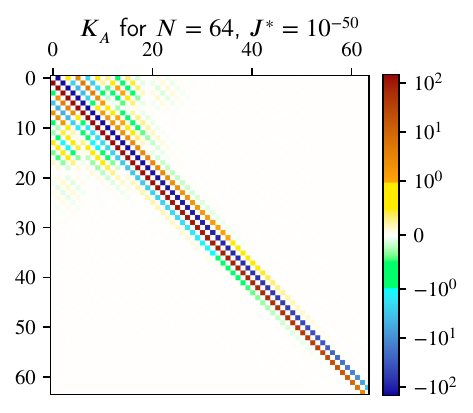}}
			\subcaptionbox{\label{fig:PlotofKSingleEdgeDefect_N64_0}}{\includegraphics[scale = 0.6]{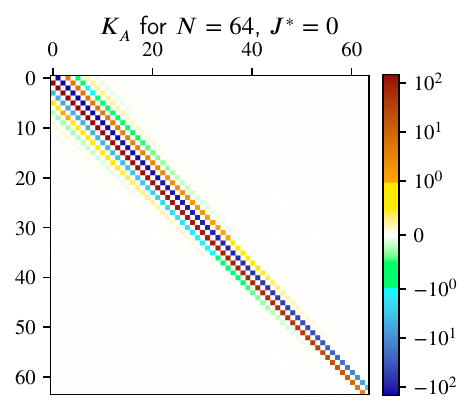}}
		\end{minipage}

		\begin{minipage}[c]{\textwidth}
			\subcaptionbox{\label{fig:SingleDefectOnBoundary_Many_NNs}}{\includegraphics[scale = 0.6]{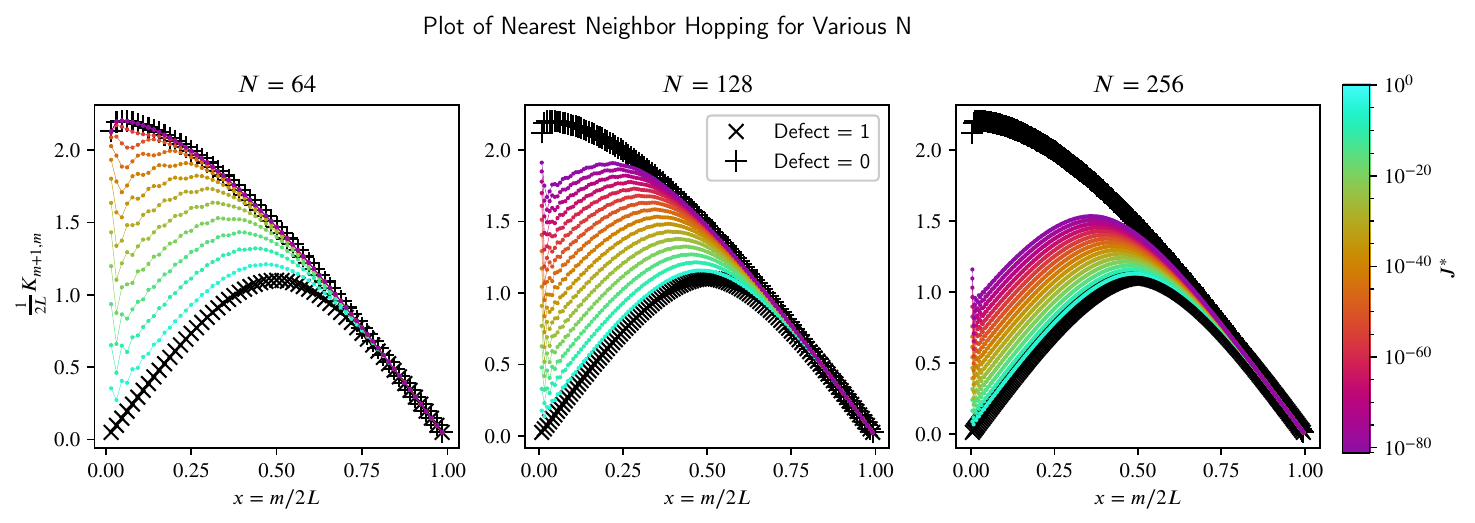}}
		\end{minipage}
		\caption{\ref{sub@fig:PlotofKSingleEdgeDefect_N64_-5}\ref{sub@fig:PlotofKSingleEdgeDefect_N64_-10}\ref{sub@fig:PlotofKSingleEdgeDefect_N64_-50}\ref{sub@fig:PlotofKSingleEdgeDefect_N64_0}) Plots of $K_A$ for various small defects, with $J^*=0$ corresponding to an open system where $A$ is one half of the system. \ref{sub@fig:SingleDefectOnBoundary_Many_NNs}) Plots of nearest neighbor interactions of $K_A$ for various system sizes and defect strengths. Note that for larger system sizes, the same coupling strength moves farther away from the open system.}\label{fig:SingleDefectOnBoundaryCluster2}
	\end{figure}

	In this section, we are interested in the case where the defect is on the boundary of the subsystem (see \ref{fig:BoundaryDefectSetup}). In particular, we want to understand how the entanglement Hamiltonian evolves as we transition from the periodic system to the open system, where both CFT solutions are known. For the open system, we send $\sin \to \cos$ and introduce a factor of $2$ in the definition of $L_1$. In \ref{fig:SingleDefectOnBoundaryCluster1}, we see that for $J^*=0.2$, the nearest neighbor interactions are relatively unchanged, with some extra couplings being prominent near the boundary with the defect. At this point, the natural question is how small do we need to make $J^*$ to come to an entanglement Hamiltonian that resembles the one for the open system. This is the focus of figure \ref{fig:SingleDefectOnBoundaryCluster2}, where see that we need very small coupling constants in order for the nearest neighbor interactions to transition, and the required strength of the defect decreases as the system size grows. 

\begin{figure}[h]	
		\centering
		\includegraphics[scale = 0.7]{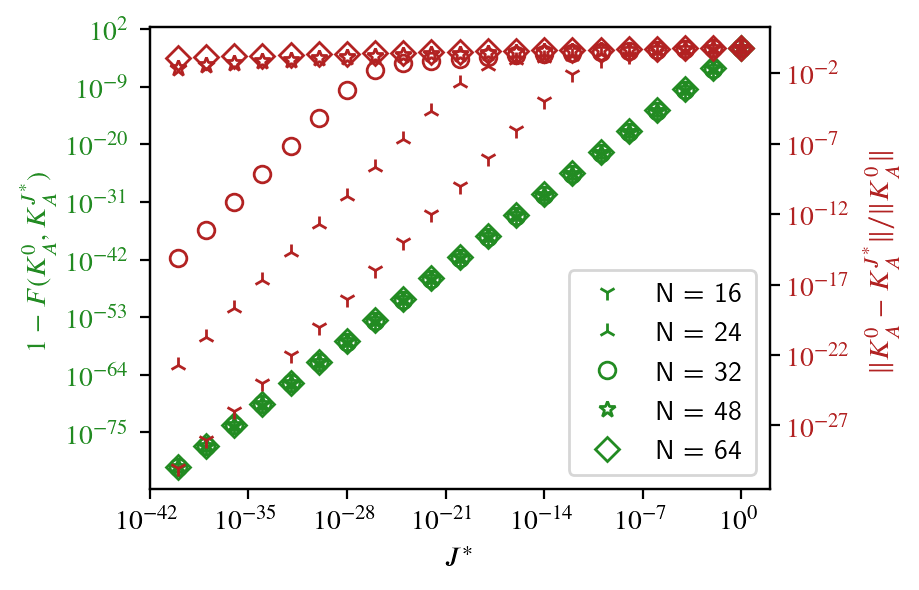}
		\caption{Infidelity of $K_A^0$ and $K_A^{J^*}$ for various defect strengths (green). This is compared to $||K_A^0-K_A^{J^*}||/||K_A^{0}||$, where $||\cdot||$ is the Frobenius norm (red). The factor of $1/||K_A^{0}||$ is there to keep the scales consistent across system sizes. The infidelity goes as $(J^*)^2$ while the norm of the difference is relatively constant up to a point, after which it goes linearly with $J^*$. The value of $J^*$ where this happens depends on the size of the system. While it is not shown in this plot, a similar change for $N=48$ happens around $J^* = 10^{-45}$ and $J^* = 10^{-60}$ for $N=64$.}\label{fig:SingleDefectOnBoundaryFidelity}
	\end{figure}

	For these small couplings, while the matrix $K_A$ itself does not resemble the open system, we can ask about the fidelity between the two reduced density matrices. Using the method presented in \cite{GuassianStateFidelity_OG,GuassianStateFidelity}\footnote{The fidelity is given by 
	\begin{equation}F(\rho_{G_1}, \rho_{G_2}) = \brac{\det\para{\frac{1-G_1}{2}}\det\para{\frac{1-G_2}{2}}}^{1/4}\brac{\det\para{1+\sqrt{\sqrt{\frac{1+G_1}{1-G_1}}\frac{1+G_2}{1-G_2}\sqrt{\frac{1+G_1}{1-G_1}}}}}^{1/2},
	\end{equation}
	where the $G_i$s are the restricted correlation matrices. Evaluating this requires high precision numerics.}, we calculate the infidelity between the open case and boundary defects along with the Frobenius norm of the difference $K_A^0-K_A^{J^*}$ and the result of this can be seen in figure \ref{fig:SingleDefectOnBoundaryFidelity}. We see that for very small couplings, both of these distance measures obey a power law dependence on the defect strength. For the Frobenius norm however, this behavior only comes into play after a critically small coupling strength which depends on the size of the system. 

	This geometry was studied numerically in \cite{Roy2022,RogersonPollmannRoy2022} and analytically in \cite{BrehmBrunner2015,Peschel_2012}, where it is discussed that the entanglement entropy in this system is given by 
	\begin{equation}
		S(L) = \frac{c_{\text{eff}}+c}{3}\log\para{\frac{N}{\pi}\sin\para{\frac{\pi L}{N}}}+S_0,
	\end{equation}
	where $L$ is the subsystem size and $S_0$ is a non-universal constant. $c_{\text{eff}}$ is the effective central charge defect on the boundary and is given by
	\begin{equation}
		c_{\text{eff}}(s) = \frac{s}{3}-\frac{1}{3}-\frac{3}{\pi^2}\brac{(s+1)\log(s+1)\log(s)+\para{s-1}\text{Li}_2(1-s)+(s+1)\text{Li}_2(-s)}
	\end{equation}
	where $s=\abs{\sin\brac{2\para{\cot^{-1}(J^*)}}}$, and $\text{Li}_2$ is the dilogarithm function. This $\cot^{-1}(J^*)$ is the parameter we found for $\omega_2$ in section \ref{sec:SingleCenteredDefect}. For $J^*=1$, $c_\text{eff}=c=0.5$, which gives the standard $S(L) = \tfrac{1}{3}\log(\eli)$ for the critical Ising model, for $J^*=0.2$, $c_{\text{eff}}\sim 0.1$ and for $J^*=10^{-5}$, $c_{\text{eff}} \sim 10^{-9}$. So even for these tiny couplings which still have effects in $K_A$, their contributions to the entropy from the universal term is negligible, and thus the difference in entropy between the open system and small $J^*$ cases will be small. This is in line with the fidelity results from above.
\section{Duality Defect}
	\begin{figure}[h]
	\centering
	\begin{tikzpicture}[scale = 0.65]
		\tikzmath{
		coordinate \c;
		coordinate \a;
		coordinate \b;
		\AngleStep = 22.5;
		\radius = 2.5;
		\spinsize = 0.1;
		\majoranasep = 5;
		\majoranasize = 0.1;
		\JOffsetSpins = 1.0;
		\JOffsetMajoranas = 1.0;
		\gWidth = 2;
		\JWidth = 2;
		\sigmazoffset = 4;
		\verticalshift = -2-\radius;
		\a = (\AngleStep-\majoranasep/2-\majoranasize*1.41cm-\AngleStep/4: {\JOffsetMajoranas*\radius});
		\b = (\AngleStep-\majoranasep/2+180-\majoranasize*1.41cm-\AngleStep/4: {\JOffsetMajoranas*\radius+0.5});
		{\draw[line width = 6, color = black!40] (\a) arc[start angle = \AngleStep-\majoranasep/2-\majoranasize*1.41cm-\AngleStep/4, end angle = \AngleStep-\majoranasep/2+180-\majoranasize*1.41cm-\AngleStep/4, radius = \radius];};
		for \x in {\AngleStep,2*\AngleStep,...,90-\AngleStep} {
			\a = (\x+\majoranasep: {\JOffsetMajoranas*\radius});
			\b = (\x-\majoranasep+\AngleStep: {\JOffsetMajoranas*\radius});
			{\draw[line width=\JWidth, JBond] (\a) -- (\b);};
		};
		for \x in {90+\AngleStep,90+2*\AngleStep,...,180} {
			\a = (\x+\majoranasep: {\JOffsetMajoranas*\radius});
			\b = (\x-\majoranasep+\AngleStep: {\JOffsetMajoranas*\radius});
			{\draw[line width=\JWidth, JBond] (\a) -- (\b);};
		};
		\a = (90-\majoranasep/2+\AngleStep-\majoranasize*1.41cm/2: {\radius});
		\b = (90-\majoranasep/2-\majoranasize*1.41cm: {\radius});
		\c = (90-\majoranasep/2-\majoranasize*1.41cm+\AngleStep/2: {0.8*\radius});
		{\draw[line width = \JWidth, color = DualityBond] (\a) .. controls (\c) .. (\b);};
		{\node[below, text width = 1.5cm] at (\c) {Duality Defect};};
		for \x in {180+\AngleStep,180+2*\AngleStep,...,360} {
			\a = (\x+\majoranasep: {\JOffsetMajoranas*\radius});
			\b = (\x-\majoranasep+\AngleStep: {\JOffsetMajoranas*\radius});
			{\draw[line width=\JWidth, JBond] (\a) -- (\b);};
		};
		\a = (\majoranasep+180: {\JOffsetMajoranas*\radius});
		\b = (0-\majoranasep+\AngleStep+180: {\JOffsetMajoranas*\radius});
		for \x in {0,\AngleStep,...,90-\AngleStep} {
			\a = (\x-\majoranasep: {\JOffsetMajoranas*\radius});
			\b = (\x+\majoranasep: {\JOffsetMajoranas*\radius});
			{\draw[line width=\gWidth, gBond] (\a) -- (\b);};
		};
		for \x in {90+\AngleStep, 90+\AngleStep*2,...,360} {
			\a = (\x-\majoranasep: {\JOffsetMajoranas*\radius});
			\b = (\x+\majoranasep: {\JOffsetMajoranas*\radius});
			{\draw[line width=\gWidth, gBond] (\a) -- (\b);};
		};
		for \x in {0,\AngleStep,...,360} {
			\c = (\x+\majoranasep: {\radius});
			{ \fill (\c) circle [radius={\majoranasize} ]; };
			\c = (\x-\majoranasep: {\radius});
			{\node[rectangle, draw, rotate = \x-\majoranasep, fill = black, minimum width=\majoranasize*1.41cm,minimum height=\majoranasize*1.41cm, inner sep=2pt] at (\c) {};};
		};
		\a = (\AngleStep-\majoranasep/2-\majoranasize*1.41cm-\AngleStep/4: {\JOffsetMajoranas*\radius+0.5});
		\b = (\AngleStep-\majoranasep/2+180-\majoranasize*1.41cm-\AngleStep/4: {\JOffsetMajoranas*\radius+0.5});
		\a = (45+\AngleStep-\majoranasep/2-\majoranasize*1.41cm-\AngleStep/4: {\JOffsetMajoranas*\radius+1});
		{\node[below] at (\a) {$A$};};
		}
	\end{tikzpicture}
	\caption{Diagram for the topological defects.}\label{fig:DualityDefect}
	\end{figure}
\begin{figure}[H]
		\centering\hspace{-4cm}
		\begin{minipage}[c]{0.8\textwidth}
			\subcaptionbox{\label{fig:Duality_NN_Ns}}{\includegraphics[scale = 0.7]{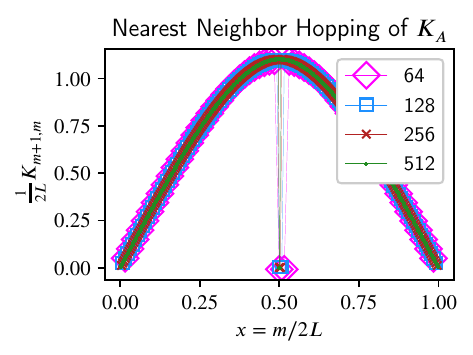}}
			\subcaptionbox{\label{fig:Duality_Skipped_Majorana}}{\includegraphics[scale = 0.7]{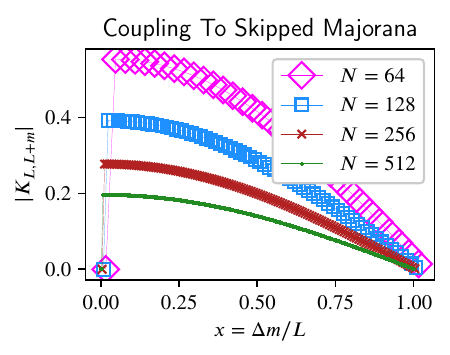}}
		\end{minipage}
		\begin{minipage}[c]{0.1\textwidth}
			\subcaptionbox{\label{fig:Duality_K}}{\includegraphics[scale = 0.7]{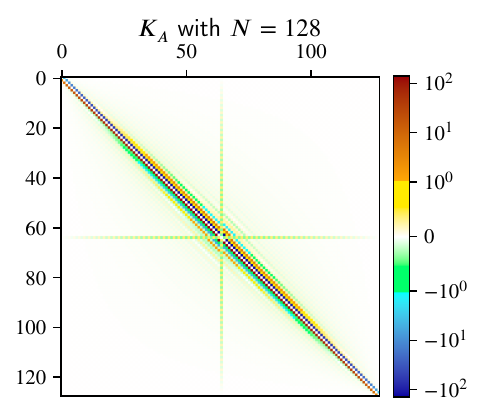}}
		\end{minipage}
		\caption{\ref{sub@fig:Duality_NN_Ns}) Nearest neighbor hopping terms of $K_A$. \ref{sub@fig:Duality_Skipped_Majorana}) The couplings to the Majorana site that is skipped by the duality defect. These are the strongest non-local couplings in the system. Note that unlike the extra couplings in the antiperiodic system, these couplings become smaller as system size increases. }\label{fig:DualityDefectPlotCluster1}
	\end{figure}
\begin{figure}[h]
		\centering
		\begin{minipage}[c]{\paperwidth}
			\hspace{0cm}
			\subcaptionbox{\label{fig:PlotOfKDualityHalfDefect}}{\includegraphics[scale = 0.6]{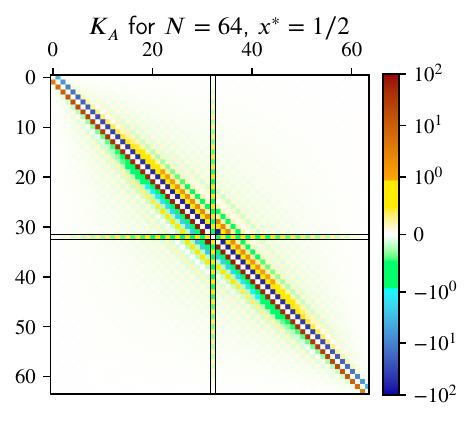}}
			\subcaptionbox{\label{fig:PlotOfKDualityOneQuarterDefect}}{\includegraphics[scale = 0.6]{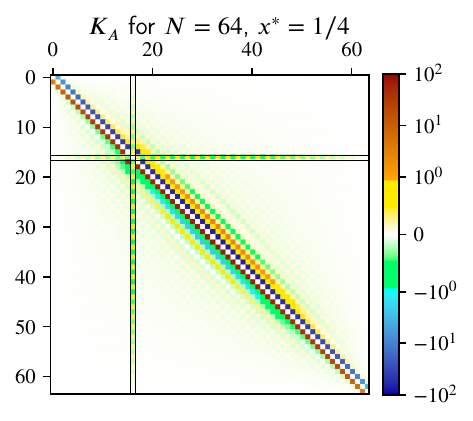}}
			\subcaptionbox{\label{fig:PlotOfKDualityOutsideDefect}}{\includegraphics[scale = 0.6]{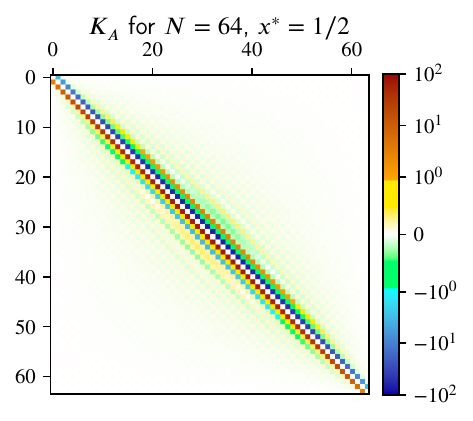}}
		\end{minipage}
		\caption{Plots of duality defect systems for different defect locations: Centered in $A$ (\ref{sub@fig:PlotOfKDualityHalfDefect}), off-centered in $A$ (\ref{sub@fig:PlotOfKDualityOneQuarterDefect}) and outside of $A$ (\ref{sub@fig:PlotOfKDualityOutsideDefect}). As with the antiperiodic defect, the location of the defect outside of $A$ does not change $K_A$. }\label{fig:DualityForDifferentPositions}
	\end{figure}

	Finally, we look at the duality defect shown in figure \ref{fig:DualityDefect}. In the Majorana language, this means skipping a single Majorana site in the chain. Starting with a centered defect, we see in figure \ref{fig:Duality_NN_Ns} that the nearest neighbor couplings follow the standard arch up to the two terms that are set to zero by the defect. In their place is an off-center bond, which is the duality defect coupling (figure \ref{fig:Duality_K}, also seen in \ref{fig:PlotOfKAntiperiodicHalfDefect}\ref{sub@fig:PlotOfKAntiperiodicOneQuarterDefect}). The vertical/horizontal elements are couplings to the skipped Majorana site generated by the zero mode. The couplings decay with system size (figure \ref{fig:Duality_Skipped_Majorana}). This is in contrast to the single defect symmetric hopping terms, which are system size independent. Also, unlike the energy defects, for which $K_A$ is checkerboard dense, the duality defect results in an entanglement Hamiltonian that is almost entirely filled, except for the self coupling term (which must be zero). In figure \ref{fig:DualityForDifferentPositions}, we see the effect of moving this duality defect around in both $A$ and in $\bar{A}$.


\bibliography{References.bib}
\bibliographystyle{iopart-num}
\end{document}